\documentclass[preprints,article,accept,moreauthors,LaTeX]{mdpi} 

\firstpage{1} 
\pubvolume{23}
\issuenum{7}
\articlenumber{879}
\pubyear{2021}
\copyrightyear{2021}
\datereceived{21 May 2021} 
\dateaccepted{07 July 2021} 
\datepublished{09 July 2021} 
\hreflink{https://doi.org/10.3390/e23070879}

\pdfoutput=1

\Title{Neural-network Quantum States for Spin-1 systems: spin-basis and parameterization effects on compactness of representations}

\TitleCitation{Neural-network Quantum States for Spin-1 systems: spin-basis and parameterization effects on compactness of representations}

\Author{Michael Y Pei $^{\dagger,*}$\orcidA{} and Stephen R Clark $^{\dagger,*}\orcidB{}$}

\AuthorNames{Michael Y Pei and Stephen R Clark}

\AuthorCitation{Pei, M.Y.; Clark, S. R.}

\address[1]{
$^{\dagger}$ \quad H.H. Wills Physics Laboratory, University of Bristol, Bristol BS8 1TL, UK}

\corres{Correspondence: michael.yuan.pei@bristol.ac.uk (M.P.), stephen.clark@bristol.ac.uk (S.R.C.)}

\usepackage{latexsym,epsfig,graphicx,bbm,bm,amssymb,wasysym}
\usepackage{amsthm,ulem}

\usepackage[export]{adjustbox}
\usepackage[framemethod=tikz]{mdframed}

\usepackage{hyperref}

\newcommand{\ket}[1]{\left | \, #1 \right \rangle}

\newcommand{\eqr}[1]{Eq.~(\ref{#1})}
\newcommand{\fir}[1]{Figure~\ref{#1}}
\newcommand{\secr}[1]{Sec.~\ref{#1}}

\newcommand{\half}{\mbox{$\textstyle \frac{1}{2}$}}
\newcommand{\bra}[1]{\left \langle #1 \, \right |}
\newcommand{\braket}[2]{\left\langle\, #1\,|\,#2\,\right\rangle}

\newcommand{\av}[1]{\langle #1\rangle}

\abstract{Neural network quantum states (NQS) have been widely applied to spin-1/2 systems where they have proven to be highly effective. The application to systems with larger on-site dimension, such as spin-1 or bosonic systems, has been explored less and predominantly using spin-1/2 Restricted Boltzmann Machines (RBMs) with a one-hot/unary encoding. Here we propose a more direct generalisation of RBMs for spin-1 that retains the key properties of the standard spin-1/2 RBM, specifically trivial product states representations, labelling freedom for the visible variables and gauge equivalence to the tensor network formulation. To test this new approach we present variational Monte Carlo (VMC) calculations for the spin-1 antiferromagnetic Heisenberg (AFH) model and benchmark it against the one-hot/unary encoded RBM demonstrating that it achieves the same accuracy with substantially fewer variational parameters. Further to this we investigate how the hidden unit complexity of NQS depend on the local single-spin basis used. Exploiting the tensor network version of our RBM we construct an analytic NQS representation of the Affleck-Kennedy-Lieb-Tasaki (AKLT) state in the $xyz$ spin-1 basis using only $M = 2N$ hidden units, compared to $M \sim O(N^2)$ required in the $S^z$ basis. Additional VMC calculations provide strong evidence that the AKLT state in fact possesses an exact {\em compact} NQS representation in the $xyz$ basis with only $M=N$ hidden units. These insights help to further unravel how to most effectively adapt the NQS framework for more complex quantum systems.}

\keyword{Neural-network Quantum States, Tensor Network Theory, Restricted Boltzmann Machines} 

\begin{document}

\section{Introduction} 

The strongly-interacting quantum many-body problem is crucial to our understanding of many intriguing physical phenomena, but also inherently difficult to treat numerically owing to the exponential growth of the Hilbert space with system size. A commonly used approximate strategy is the variational method where a trial state, characterised by a tractable number of variational parameters, is optimised in energy. The effectiveness of this approach is highly dependent on the ansatz having an expressive form that can be systematically improved, to minimise bias, while also allowing relevant observables to be evaluated efficiently. Tensor networks have provided several examples of such ansatzes, with matrix product states (MPS)~\cite{verstraete08,orus14} displaying impressive accuracy in one-dimensional systems, along with projected entangled-pair states (PEPS)~\cite{verstraete08,cirac_mps10}, tree tensor networks (TTN)~\cite{shi2006,murg10} and multi-scale entanglement renormalization ansatz (MERA)~\cite{evenbly13} making two-dimensional systems accessible. Recently artificial neural networks (ANNs) have emerged as another class of highly flexible variational ansatzes with many variants such as restricted Boltzmann machines (RBM)~\cite{carleo_nqs17}, Deep Boltzmann Machines (DBM)~\cite{gao_dbm17,carleo_dbm18,he_mdbm19}, convolutional neural networks (CNN)~\cite{choo_fmag19,naoki2020,markus2020,liang2021,levine19} and feed-forward neural networks (FFNN)~\cite{saito18,choo18,luo19,adams2020}. An important advantage of ANNs is that they are highly flexible and can be applied to any number of spatial dimensions making them a powerful method for tackling the subtle physics seen in two-dimensional systems.
\par

Although one of the simplest ANN variants RBMs have seen widespread applications including for open quantum systems~\cite{torlai18,vicentini19,hartmann19,nobuyuki19}, frustrated spin problems~\cite{westerhout20,nomura2021}, quantum circuit simulation~\cite{jonsson18,freitas18,bausch20} and more. There are several reasons for their continued use. First, their simple structure allows for efficient sampling crucial for applying variational Monte Carlo (VMC)~\cite{gubernatis16,becca17}. Second, RBMs are also a good candidate for a weakly biased ansatz, given that they are capable of exactly representing arbitrary states when their hidden unit number $M$ scales exponentially with system size $N$. Third, RBMs are capable of representing states with volume-law entanglement~\cite{deng17}, which further distinguishes them from tensor networks~\cite{eisert10} despite their conceptual similarities~\cite{clark_cps18,chen18,collura20}. Finally, there are also numerous classes of states with efficient exact RBM representations, including graph states~\cite{gao_dbm17}, spin Jastrow states such as Laughlin states~\cite{glasser18,kaubruegger18,clark_cps18}, general stabiliser states such as the toric code~\cite{zheng_stbl19,zhang_rbm2stbl18,lu19,jia_surf19}, as well as more exotic hypergraph states and XS-stabilizer states~\cite{lu19}. Recently we found that all but the last class listed in fact have RBM representations requiring $M<N$ hidden units~\cite{Pei21}, illustrating how even very modestly sized RBMs have significant representational power. 
\par

Despite their efficacy for spin-$\half$ systems, the application of RBMs to systems with a local on-site dimension $d > 2$, such as spin-1 or bosonic systems, has been limited with convolutional or feedforward neural networks generally being favoured~\cite{saito17,saito18,choo18}. The typical approach in machine learning to handle models with multinomial or categorical variables is so-called ``one-hot" or ``unary" encoding~\cite{guo16}. Rather than representing a physical degree of freedom directly with one visible unit this approach encodes the possible local physical states into a set of binary visible units. While this approach leverages the power of binary or spin-$\half$ RBMs it multiplies the number of visible units by a factor $d$, significantly increasing the parameter count and complexity of the optimisation. As a consequence, studies utilising unary encoding so far, for example on the Bose Hubbard model~\cite{mcbrian19,vargas20}, have been limited to small system sizes. Thus there is need to devise more efficient RBM constructions tailored for $d>2$ systems. 
\par

Progress has been made in this direction in recent work~\cite{vieijra2020} where multivalued RBMs were applied directly to the 
one-dimensional spin-1 antiferromagnetic Heisenberg model (AFH) and substantially enhanced by incorporating a transformation to a coupled SU(2) symmetric basis. Complementary to this here we propose and study a direct generalisation of the RBMs\footnote{As we will not be examining other network architectures, we will use RBM and NQS interchangeably in this paper.} to spin-1 systems that retains key properties of spin-$\half$ RBM with a minimal increase in variational parameters. Specifically, the ability to describe arbitrary product states without hidden units, invariance of the parameterisation to the values assigned to visible variables (labelling freedom), and equivalence to the tensor network formulation. This leads to the introduction of new quadratic bias and interaction weights in the RBM effective energy function. We demonstrate the effectiveness of the new formulation via VMC calculations for spin-1 AFH model where it is seen to deliver the same accuracy as unary encoding but with substantially fewer variational parameters. Additionally we also investigate how the local single-spin basis affects the hidden unit complexity of a state by performing calculations in both the $S^{z}$ and $xyz$ spin-1 bases. For the AFH model with periodic boundary conditions we find that the $S^{z}$ is more accurate. A useful advantage of our new spin-1 RBM formulation is that it permits tensor network based analytic constructions. Focusing on the paradigmatic Affleck-Kennedy-Lieb-Tasaki (AKLT) model we give explicit exact NQS representations in both the $S^{z}$ and $xyz$ bases. Our $S^z$ basis NQS construction displays the expected~\cite{glasser18} $M \sim O(N^2)$ scaling, while the simplification of the amplitude structure in the $xyz$ basis gives an NQS construction with $M = 2N$ hidden units. By using VMC calculations we find compelling evidence that the AKLT state in fact only requires $M=N$ hidden units to be represented exactly in the $xyz$ basis. 
\par

The structure of this paper is as follows. We briefly outline the VMC method applied to the many body problem in \secr{sec:vmc} and discuss desirable properties for variational ansatzes. Next in \secr{sec:nqs} we discuss the RBM in its spin-$\half$ form and analyse its key properties. In \secr{sec:spin1} we introduce a new generalisation of NQS to spin-1 systems designed to mimic these properties and present VMC results for the AFH model. We then introduce analytic constructions for the AKLT model in \secr{sec:aklt} followed by VMC calculations in both the $S^z$ and $xyz$ basis. Finally, in \secr{sec:conclusion} we conclude and discuss some open problems.
\par

\section{Variational Monte Carlo and the many-body problem} \label{sec:vmc}

\subsection{Quantum many-body problem}
In this work we will focus on physical systems composed of $N$ localised spinful particles. Each particle is described by a vector of spin operators $\hat{\bf S}_j = (\hat{S}^x_j,\hat{S}^y_j,\hat{S}^z_j)$ for $j=1,2,\dots,N$. Typically the eigenstates $\ket{S_j}$ of $\hat{S}^z_j$ are used as the local spin basis and from which the $S^z$ basis for the full system is constructed as $\ket{\bm S} =\ket{S_1}\otimes \cdots \otimes \ket{S_N}$ where ${\bm S} = (S_1,S_2,\dots,S_N)$. Any many-body quantum state for these $N$ spins can then be expanded in this basis as
\begin{equation}
\ket{\Psi} = \sum_{\bm S} \Psi({\bm S}) \ket{\bm S},
\end{equation}
via its complex amplitudes $\Psi({\bm S})$. In the spin-$\half$ case we have $\hat{S}^\alpha_j = \half\hat{\sigma}^\alpha_j$ (taking $\hbar =1$ throughout) for $\alpha =\{x,y,z\}$ defined by the Pauli matrices, and $S_j \in \{+\half \equiv\, \uparrow,-\half\equiv\,\downarrow\}$. In the spin-1 case we have
\begin{eqnarray*}
\qquad \hat{S}^x_j &=& \frac{1}{\sqrt{2}}\left[
\begin{array}{ccc}
0  & 1  & 0  \\
1  & 0  & 1  \\
0  & 1  & 0  
\end{array}
\right], \quad 
\hat{S}^y_j = \frac{1}{\sqrt{2}}\left[
\begin{array}{ccc}
0  & -{\rm i}  & 0  \\
{\rm i}  &  0 & -{\rm i}  \\
0  & {\rm i}  & 0  
\end{array}
\right], \quad
\hat{S}^z_j = \left[
\begin{array}{ccc}
1  & 0  & 0  \\
0  & 0  & 0  \\
0  & 0  & -1  
\end{array}
\right],
\end{eqnarray*}
with $S_j \in \{+1 \equiv\, \Uparrow,0,-1\equiv\,\Downarrow\}$.
\par

A key challenge in many-body physics is to find the ground state $\ket{\Psi_0}$ of a system governed by an interacting Hamiltonian $\hat{H}$. In the context of spin systems $\hat{H}$ will include terms that are products of the spin operators $\hat{\bf S}_j$ over two or more spins. Given that the expectation value of any observable $\hat{A}$ for a general (unnormalised) state $\ket{\Psi}$ is 
\begin{equation}
\av{A}_\Psi = \frac{\bra{\Psi}\hat{A}\ket{\Psi}}{\braket{\Psi}{\Psi}},
\end{equation}
the variational approach reformulates the eigenvalue problem $\hat{H}\ket{\Psi_0} = E_0\ket{\Psi_0}$ as the minimisation of the energy $E = \av{\hat{H}}_\Psi$ over the exponentially many amplitudes $\Psi({\bm S})$. Performing this task exactly is only feasible for small systems $N \sim O(10)$~\cite{lauchli12}. 
\par
\subsection{Variational Monte Carlo method}
A way to circumvent the ``curse of dimensionality" is to instead to restrict the optimisation over a specialised class of states $\ket{\Psi_{\bm p}}$, dependent on parameters $\bm p$ whose number $n_{\rm params}$ scales polynomially with $N$. The variational principle $E_0 \leq E_{{\bm p}_0} = {\rm min}_{\bm p} \av{\hat{H}}_{\Psi_{\bm p}}$ then provides a route to finding the best approximation $\ket{\Psi_{{\bm p}_0}}$ within the ansatz for $\ket{\Psi_0}$. The flexibility and utility of variational ansatz are greatly enhanced if instead of computing expectation values $\av{A}_{\Psi_{\bm p}}$ exactly we evaluate them approximately using Monte Carlo sampling. This approach is called {\em variational Monte Carlo} (VMC) and is described in detail in App.~\ref{sec:samp}. A key feature of VMC is that only ratios of amplitudes for an ansatz $\Psi_{\bm p}({\bm S})/\Psi_{\bm p}({\bm S}')$ between different spin states $\ket{\bm S}$ and $\ket{{\bm S}'}$ are needed within the algorithm\footnote{For this reason we will ignore the normalisation of quantum states in this work.}. To be efficient we thus require that these amplitude ratios for our ansatz can be evaluated with an effort scaling polynomially with $N$. 
\par

All numerical calculations presented in this paper employ the powerful stochastic reconfiguration method~\cite{sorella01,becca17} for optimising the parameters $\bm p$. While ansatze with a larger number of parameters may describe more varied amplitude structure, in principle allowing greater accuracy, the effort to optimise them can scale as $O(n_{\rm params}^3)$. It is therefore crucial that the functional form of any candidate ansatz $\ket{\Psi_{\bm p}}$ is judicious in its use of variational parameters to avoid excessive redundancies.  
\par

\section{Neural-network quantum states in spin-1/2} \label{sec:nqs}
In this section we review NQS in terms of RBMs, discussing their form from both an energy function and tensor network perspective. In doing so we highlight key properties of NQS that are desirable for expressiveness and the ability to represent important families of states.

\subsection{Restricted Boltzmann machine approach} \label{sec:rbm}
Restricted Boltzmann machines consist of two sets of units, $N$ {\em visible} units representing the physical system, and $M$ {\em hidden} units to be marginalised out. The units are characterised by a Boltzmann-like combined probability distribution $p({\bm v},{\bm h}) = \exp(-\mathcal{E}_{\bm \lambda})$ with an effective ``energy" function
\begin{equation}
\mathcal{E}_{\bm \lambda}({\bm v},{\bm h}) = - \sum_{j=1}^N a_j v_j - \sum_{i=1}^M b_i h_i - \sum_{i=1}^M\sum_{j=1}^N w_{ij}h_i v_j, \label{rbm_energy}
\end{equation}
where ${\bm \lambda} = \{{\bm a},{\bm b},{\bm w}\}$ is the set of $MN + M + N$ complex parameters consisting of $N$ visible biases ${\bm a} = (a_{1},\cdots,a_{N})$, $M$ hidden biases ${\bm b} = (b_{1},\cdots,b_{M})$ and $M \times N$ hidden-visible interaction weights ${\bm w} = [w_{11},\cdots,w_{1N},\cdots,w_{MN}]$. While $p({\bm v},{\bm h})$ is Boltzmann once the hidden units are traced out we are left with a more complex marginal distribution for the visible units from which the NQS amplitudes are derived~\cite{carleo_nqs17}
\begin{equation}
\Psi_{\bm \lambda}({\bm S} \mapsto {\bm v}) = \sum_{{\bm h}} \exp\left[ -\mathcal{E}_{\bm \lambda}({\bm v},{\bm h})\right]. \label{rbm}
\end{equation}
The RBM architecture is shown diagrammatically in \fir{fig:rbm}. 

\begin{figure}[ht]
\begin{center}
\includegraphics[scale=0.5]{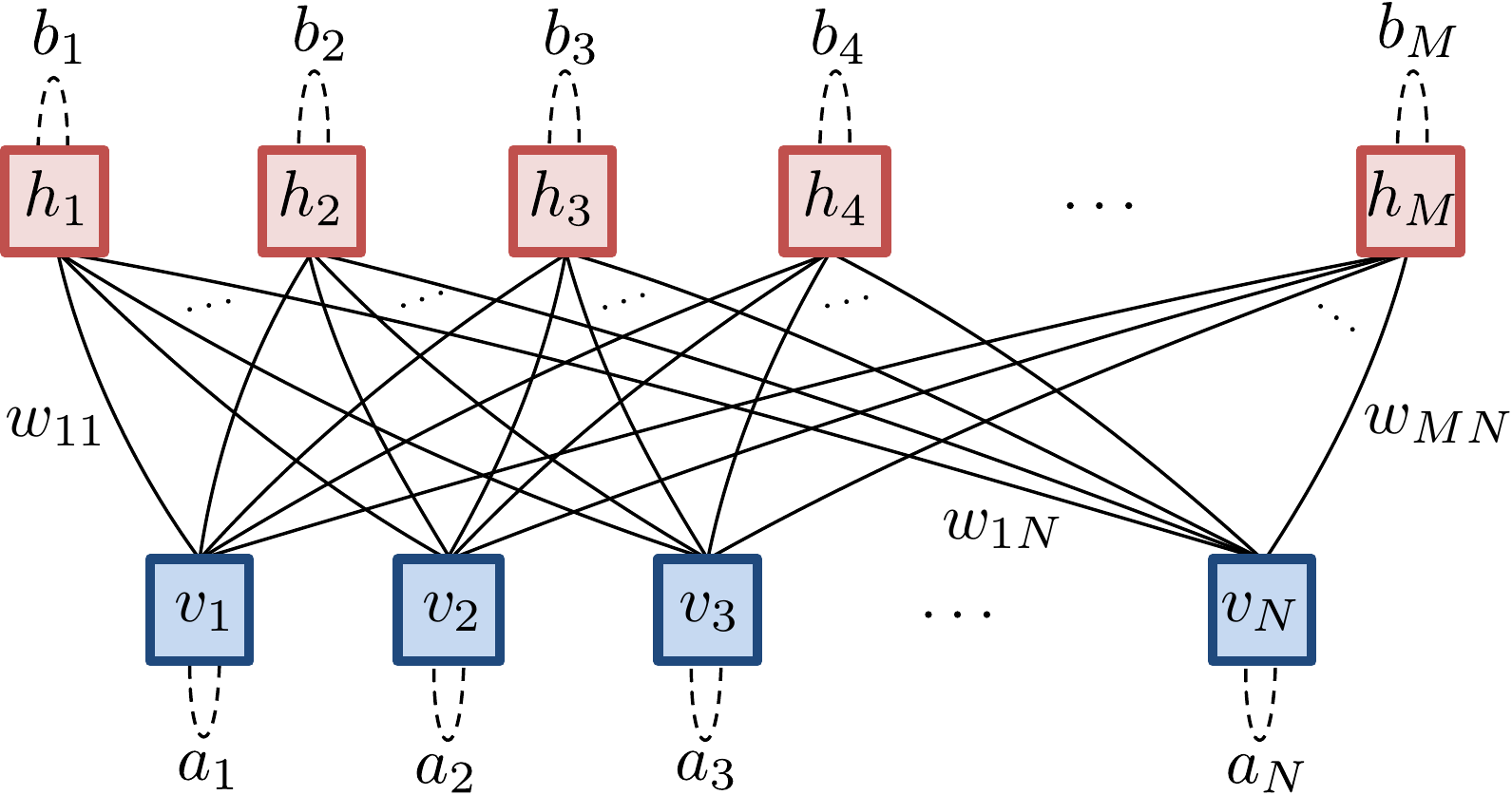}
\end{center}
\caption{The bipartite graph of an RBM depicting interaction weights $\bm w$ as edges shown as solid arcs between hidden and visible units. For completeness the biases $\bm a$ and $\bm b$ on each unit are depicted here as self-loop edges with dotted arcs to distinguish them from the interactions. The diagram here has been duplicated from Ref.~\cite{Pei21} and is presented here again for clarity.}
\label{fig:rbm}
\end{figure}

Typically visible $v_j$ and hidden $h_i$ unit variables are taken as two-valued. The pair of unique values $\{\mu,\nu\}$ taken by units within the energy function \eqr{rbm_energy} can be freely chosen, and need not coincide with the eigenvalues $S_j \in \{+\half,-\half\}$ of the local operator $\hat{S}^z$ chosen to define the physical basis. In \eqr{rbm} we emphasise this {\em labelling freedom} by explicitly introducing a mapping between physical and visible configuration\footnote{Commonly with NQS an implicit choice ${\bm v} = {\bm S}$ is made nullifying the need for this distinction, but it will be useful here.} ${\bm S} \mapsto {\bm v}$. Canonical choices $v_j$ are number-like $\{0,1\}$ or Ising-like $\{+1,-1\}$ values. We will adopt Ising-like visible variables, in which case it follows that an arbitrary visible unit taking values
\begin{equation}
S_j = \left\{
\begin{array}{ccr}
\uparrow  & \mapsto & \mu    \\
\downarrow  & \mapsto & \nu  \\    
\end{array}\right\} = \tilde{v}_j,
\end{equation}
is generated by a shift and rescaling of $v_j \in \{+1,-1\}$ as $\tilde{v}_j = \half(\mu+\nu) + \half(\mu - \nu)v_j$. Since this is a linear transformation it is entirely accommodated by redefining the energy function $\mathcal{E}_{\bm \lambda}({\bm v},{\bm h})$ parameters $\bm \lambda$ without changing its functional form. We shall see shortly that labelling freedom is a crucial ingredient for generalising RBMs to spin-1 system using higher-dimensional visible units in \secr{sec:spin1}. Tracing out Ising-like hidden variables results in the amplitude expansion~\cite{carleo_nqs17}
\begin{eqnarray}
\Psi_{\bm \lambda}({\bm S} \mapsto {\bm v}) &=& \prod_{j=1}^N e^{a_jv_j}  \prod_{i=1}^M 2\cosh\left(b_i + \sum_{j=1}^N w_{ij}v_j\right), \label{rbm_amps}
\end{eqnarray}
that is commonly used and numerically convenient.

A key property of spin-$\half$ RBMs is that they can represent any generic product state of the form
\begin{equation}
    \left(c_{+}^{(1)}\ket{\uparrow} + c_{-}^{(1)}\ket{\downarrow}\right) \otimes \cdots \otimes \left(c_{+}^{(N)}\ket{\uparrow} + c_{-}^{(N)}\ket{\downarrow}\right), \label{eq:prod_state}
\end{equation}
without hidden units by simply setting their visible biases to 
\begin{equation}
    a_{j} = \frac{\ln\left(c^{(j)}_{+}\right) - \ln\left(c^{(j)}_{-}\right)}{\mu - \nu}.
\end{equation}
Hidden units are thus necessary to describe entangled states, however unlike tensor networks, there is no direct quantitative relation between them. Indeed Ref.~\cite{deng17} reports that for random NQS adding hidden units actually decreases the amount of entanglement in a state. Nevertheless, increasing the number $M$ of hidden units increases the expressiveness of the NQS allowing more complex correlations within $\Psi_{\bm \lambda}({\bm S} \mapsto {\bm v})$ to be encoded. Formally the NQS ansatz can represent any arbitrary state exactly in the limit of $M \rightarrow 2^N$ \cite{leroux08}. More usefully there are important classes of states that possess highly accurate approximate or even exact NQS representations with an efficient scaling $M \sim {\rm poly}(N)$ ~\cite{glasser18,clark_cps18,lu19,jia_surf19,zhang_rbm2stbl18,zheng_stbl19}. A particularly tractable NQS hidden unit complexity is typified by the following:
\theoremstyle{definition}
\begin{Definition}[Compact NQS]
States with an exact NQS representation where $M \leq N$ will be denoted as compact.
\end{Definition}
\noindent As we demonstrated in Ref.~\cite{Pei21} important classes of state including Jastrow, graph and stabiliser states all have compact NQS representations.

\subsection{Tensor network approach} \label{sec:tnt}

\begin{figure}[ht]
\begin{center}
\includegraphics[scale=0.5]{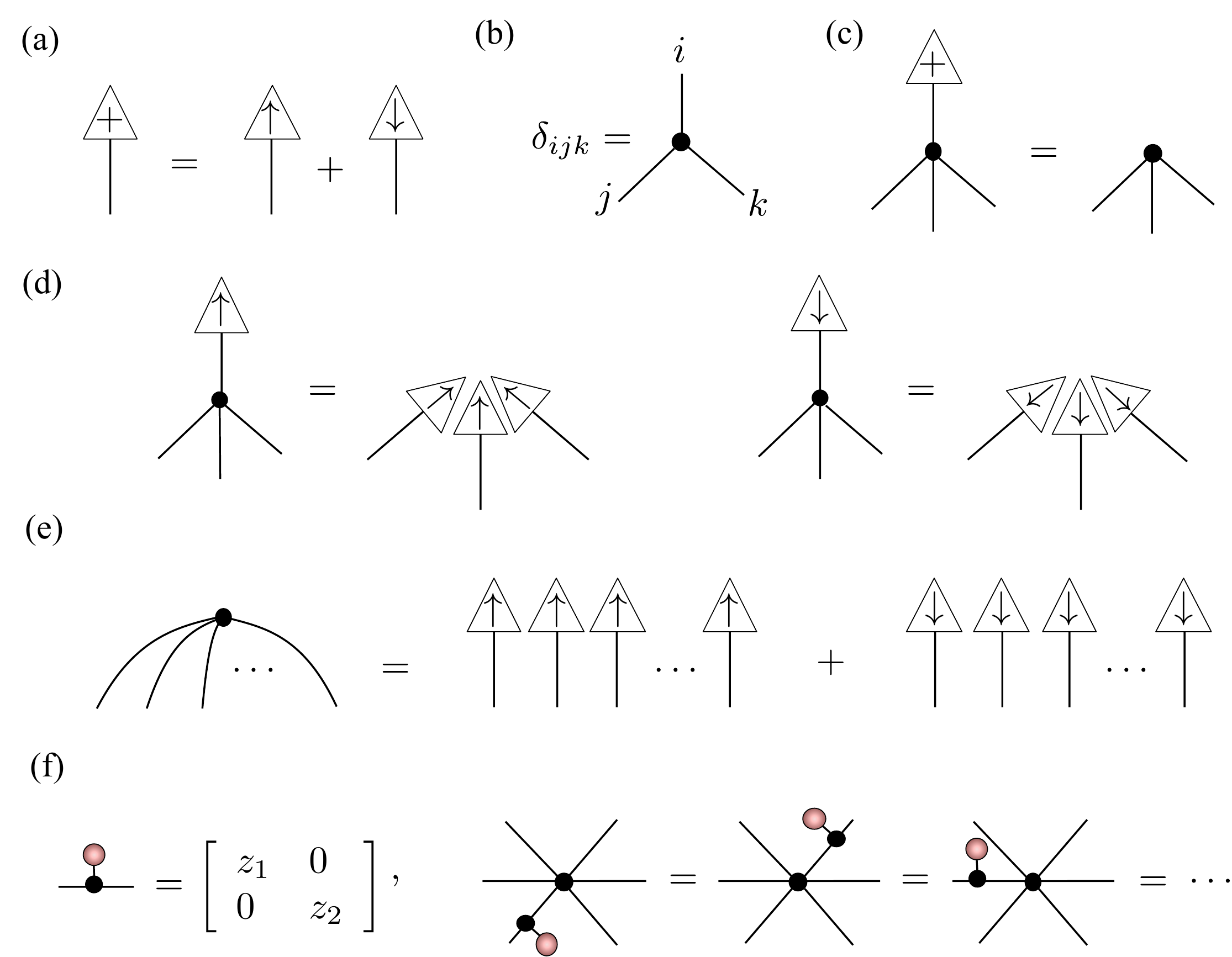}
\end{center}
\caption{A visual summary of the properties of the two-dimensional COPY tensor, showing diagrammatically the basic operations needed to construct the NQS tensor network and expose its gauge freedom~\cite{Pei21}.}
\label{fig:copy2d}
\end{figure}

An alternative formulation of NQS views them as a tensor network. From this perspective the amplitudes $\Psi({\bm S})$ are recast as elements of an order-$N$ tensor $\Psi_{S_1S_2\cdots S_N}$, and this structureless tensor can be decomposed into a set of lower order tensors contracted together~\cite{verstraete08,orus14}. Here we will make repeated use of tensor network diagrams that form an important analytical tool. Here generic tensors of any order are represented as shapes, most often a circle $\circ$ colour shaded to guide the eye, with protruding legs for each index they possess. Contraction of tensors is then represented by the joining of legs via graphical equations like this
\begin{equation}
\includegraphics[scale=0.5,valign=c]{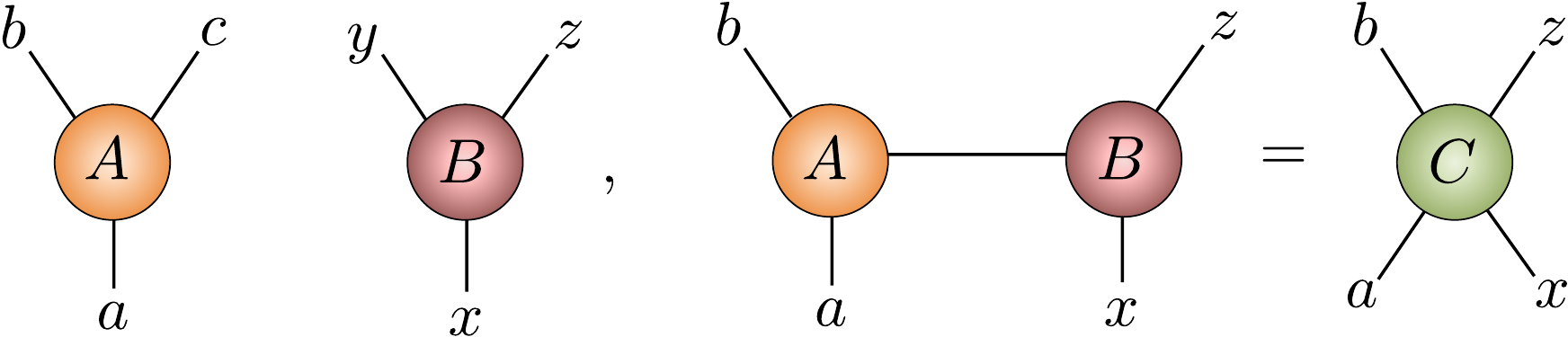} ,
\end{equation}
depicting $C_{abxz} = \sum_\alpha A_{ab\alpha}B_{x\alpha z}$. When there is a symbol inside a shape, it represents tensors with fixed elements or some specific structure. For example the order-1 tensors with $\uparrow$ or $\downarrow$ symbols are the representation of the spin-$\half$ $S^z$ basis states $\ket{\uparrow} = ({1 \atop 0})$ and $\ket{\downarrow} = ({0 \atop 1})$, while the tensor shown in \fir{fig:copy2d}(a) represents $\ket{+} = \ket{\uparrow} + \ket{\downarrow}$. A special tensor we will use frequently is the COPY tensor~\cite{biamonte11,clark_cps18}, denoted by a dot $\bullet$, which is the multidimensional generalisation of the $2\times 2$ identity matrix. Its order-3 variant is shown in \fir{fig:copy2d}(b). It possesses a number of important properties. First, contracting any COPY tensor leg with $\ket{+}$ reduces the order of the COPY tensor by 1, as shown in \fir{fig:copy2d}(c). Second, contracting any COPY tensor leg with an $S^z$ basis state factorises it by duplicating the basis vector across all legs, as shown in \fir{fig:copy2d}(d), motivating its name. This property makes COPY tensors a useful glue for constructing sampleable tensor networks. An order-$N$ COPY tensor can itself be viewed as a representation of a GHZ state, shown in \fir{fig:copy2d}(e). Finally, \fir{fig:copy2d}(f) shows a generic diagonal matrix attached to a leg of a COPY tensor can be commuted across to any of the COPY dot's other legs.

With these concepts in place, the bipartite graph structure of RBM graph readily translates to a tensor network
\begin{equation}
\includegraphics[scale=0.5,valign=c]{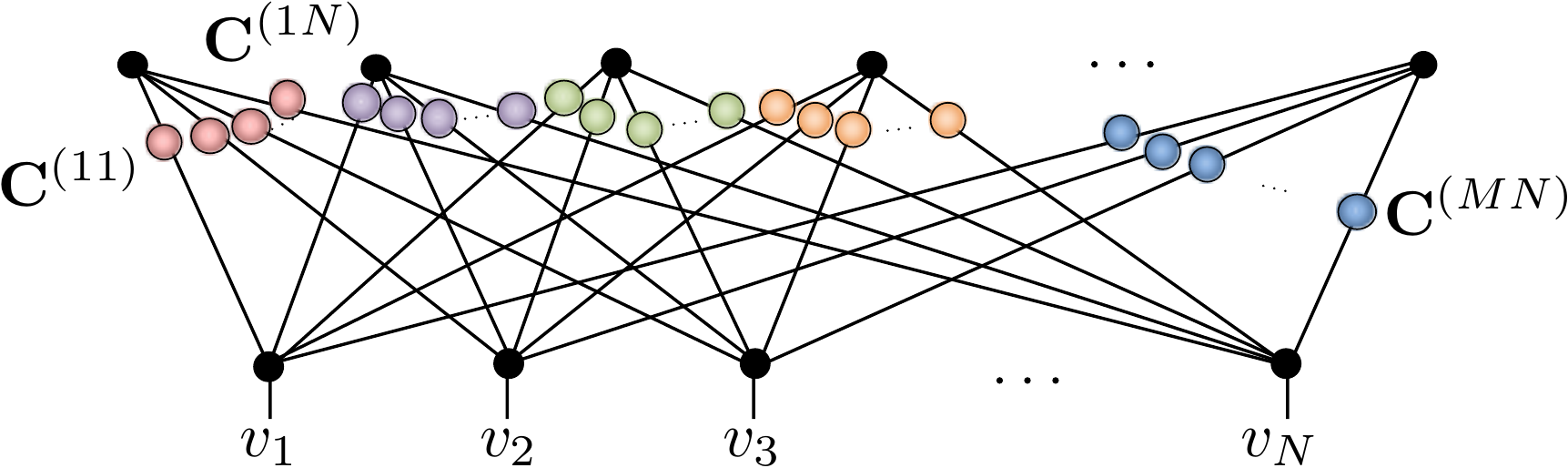} , \label{eq:nqs_tn}
\end{equation}
in which the vertices representing the visible and hidden units are replaced by COPY tensors, and order-2 tensors, or $2 \times 2$ coupling matrices ${\bm C}^{(ij)}$, are introduced on each edge of the graph. Based on \fir{fig:copy2d}(e) we can thus view each hidden unit in an NQS as contributing within an amplitude-wise product a locally deformed GHZ state. The NQS tensor network readily displays the key properties of the spin-$\half$ RBM. They can trivially represent product states in \eqr{eq:prod_state} by using one hidden unit with rank-1 coupling matrices ${\bm P}$ containing the coefficients $\{c^{(j)}_{+}, c^{(j)}_{-}\}$ that subsequently factorises out as
\begin{equation}
\includegraphics[scale=0.5,valign=c]{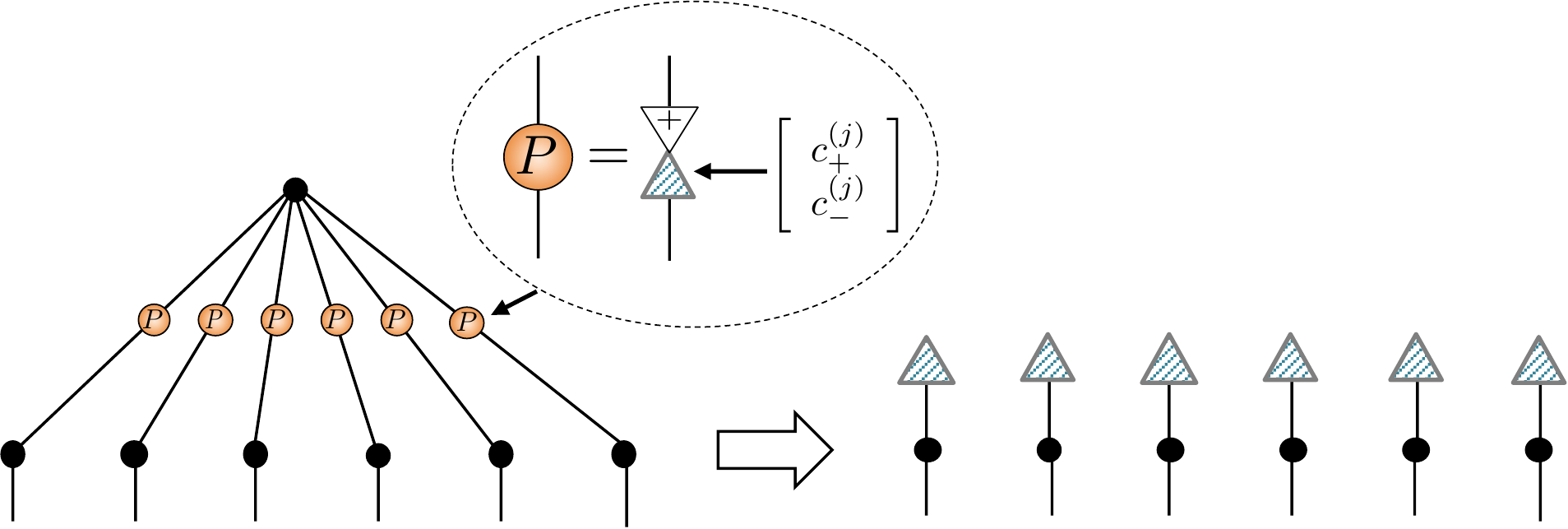} . \label{eq:nqs_prod_state}
\end{equation}
Moreover since the elements of coupling matrices ${\bm C}^{(ij)}$ have no functional dependence on any visible variables they manifestly display labelling freedom. Crucially, while the NQS tensor network appears to have $4MN$ variational parameters it exhibits significant gauge freedom due to the ability to reshuffle diagonal matrices across COPY tensors, as shown in \fir{fig:copy2d}(g). Consequently most of the elements in the coupling matrices can be extracted and combined at the COPY tensors reducing the number of independent parameters to $N + M + NM$, identical to the RBM formulation~\cite{Pei21}. This equivalence allows either the NQS tensor network or RBM approach to be used interchangeably, and any direct generalisation of RBMs to spin-1 should retain this feature. 
 
\section{Generalisation to spin-1 systems} \label{sec:spin1}
A tentative definition of a spin-1 RBM can be made by simply using the three-valued visible variables $v_j$ with a natural Ising-like physical to visible mapping
\begin{equation}
S_j = \left\{
\begin{array}{ccr}
\Uparrow  & \mapsto & +1    \\
0  & \mapsto & 0  \\
\Downarrow & \mapsto & -1    
\end{array}\right\} = v_j,
\end{equation}
in the standard linear energy function \eqr{rbm_energy}. By retaining two-valued Ising-like hidden units $h_i \in \{+1,-1\}$ the amplitudes $\Psi_{\bm \lambda}({\bm S} \mapsto {\bm v}) = \sum_{{\bm h}} \exp\left[\mathcal{E}_{\bm \lambda}({\bm v},{\bm h})\right]$ continue to be given by \eqr{rbm_amps}, but now admitting three-valued visible variables. 
\par
This spin-1 generalisation of the RBM lacks the key properties of the spin-$\half$ variant. First, $\Psi_{\bm \lambda}({\bm S} \mapsto {\bm v})$ cannot easily describe a generic product spin state  
\begin{equation}
\qquad\left(c^{(1)}_{+}\ket{\Uparrow} + c^{(1)}_0\ket{0} + c^{(1)}_{-}\ket{\Downarrow}\right)\otimes \cdots \otimes \left(c^{(N)}_{+}\ket{\Uparrow} + c^{(N)}_0\ket{0} + c^{(N)}_-\ket{\Downarrow}\right), \label{eq:spin1_prod_state}
\end{equation}
with coefficients $c^{(j)}_\alpha$, since the visible bias term $\exp(a_jv_j)$ cannot discriminate each of the local spin-1 states. This points to issues of expressiveness since we are using the same set of $MN + M + N$ parameters $\{{\bm a},{\bm b},{\bm w}\}$ to describe a bigger state space. Second, the energy function \eqr{rbm_energy} does not possess labelling freedom for the visible units values. An arbitrary physical to visible mapping
\begin{equation}
S_j = \left\{
\begin{array}{ccr}
\Uparrow  & \mapsto & \mu    \\
0  & \mapsto & \nu  \\
\Downarrow & \mapsto & \omega    
\end{array}\right\} = \tilde{v}_j,
\end{equation}
is generated from an Ising-like variable $v_j = \{+1,0,-1\}$ via the quadratic transformation $\tilde{v}_j = \nu + \half(\mu - \omega)v_j + \half(\mu+\omega - 2\nu)v_j^2$. Consequently transforming visible variables in this way cannot be accommodated in $\mathcal{E}_{\bm \lambda}({\bm v},{\bm h})$ by changing only the parameters $\{{\bm a},{\bm b},{\bm w}\}$. One way to avoid these issues is to use ``one-hot" or ``unary" encoding which has been successfully applied to both spin-1~\cite{lu19} and bosonic systems~\cite{mcbrian19,vargas20}.

\subsection{Unary encoding approach} \label{sec:rbm_unary}
Unary encoding applies the spin-$\half$ RBM formalism of \eqr{rbm} to a larger state space of each spin-1 by mapping the physical system into one comprising a larger number of spin-$\half$ particles. Specifically, the local $S^z$ basis of each spin-1 is mapped on to three spin-$\half$'s as
\begin{equation}
\ket{\Uparrow} = \ket{\uparrow\uparrow\downarrow}, \quad \ket{0} = \ket{\uparrow\downarrow\uparrow}, \quad \ket{\Downarrow} = \ket{\downarrow\uparrow\uparrow}.
\end{equation}
Physical states of the spin-1 system are now contained in the unary encoded subspace where only a single-excitation occurs within any three-site unit cell which facilitates the efficient projection of the representation. This naturally generalises to a $d$-dimensional local state space.
\par
By using this mapping a spin-$\half$ RBM can be applied inheriting its labelling freedom. Owing to the single-excitation projection product states in \eqr{eq:spin1_prod_state} are readily described by the visible biases $a_{j{\rm a}}, a_{j{\rm b}}$ and $a_{j{\rm c}}$ associated with the three spin-$\half$'s \{a,b,c\} encoding a given spin-1. In general for a $d$-dimensional local state space unary encoding accounts for the enlarged state space by increasing the number of variational parameters to $n_{\rm params} = M + dN + dNM$ compared to the naive spin-1 RBM. However, there are some obvious deficiencies of this approach. First, unary encoding appears to unnecessarily enlarge the parameter count, as evidenced by the fact that it increases it even for $d=2$. This will significantly increase the computational cost. Second, splitting the interaction weights $w$ across a unary cell makes the interpretation of any individual hidden unit's contribution to the physical state rather opaque.
\par

\subsection{Defining a spin-1 RBM and tensor network} \label{sec:spin1rbm}
\par
These deficiencies show that a more general RBM energy function is needed that is sensitive to the three-valued nature of the visible units through the inclusion of terms involving the square of visible variables\footnote{A similar approach is likely to have been used in Ref.~\cite{vieijra2020} already, although it was not explicitly stated.}. This motivates the following definition:
\theoremstyle{definition}
\begin{Definition}[spin-1 NQS]
We introduce a direct spin-1 RBM as the ansatz with amplitudes $\Psi_{\bm \Lambda}({\bm S}\mapsto{\bm v}) = \sum_{{\bm h}} \exp\left[\mathcal{E}_{\bm \Lambda}({\bm v},{\bm h})\right]$ via the energy function
\begin{eqnarray}
\quad \mathcal{E}_{\bm \Lambda}({\bm v},{\bm h}) = \sum_{j=1}^N a_j v_j +  \sum_{j=1}^NA_j v_j^2 + \sum_{i=1}^M b_i h_i + \sum_{i=1}^M\sum_{j=1}^N w_{ij}h_iv_j + \sum_{i=1}^M\sum_{j=1}^N W_{ij}h_iv^2_j,\quad \label{rbm_energy_s1}
\end{eqnarray}
defined by the parameters ${\bm \Lambda} = \{{\bm a},{\bm b},{\bm w},{\bm W},{\bm A}\}$.
\end{Definition}
The new contributions to a spin-1 RBM are $\bm W$, an $M \times N$-dimensional matrix of quadratic interactions, and $\bm A$, an $N$-dimensional vector of quadratic visible biases. There are now $2MN + 2N + M$ complex parameters in total. Tracing out the two-valued Ising-like hidden units gives amplitudes
\begin{eqnarray}
\qquad \Psi_{\bm \Lambda}({\bm S}\mapsto{\bm v}) &=& \prod_{i=1}^N e^{a_iv_i + A_iv^2_i } \prod_{j=1}^M 2\cosh\left(b_j + \sum_{i=1}^N w_{ij}v_i+ \sum_{i=1}^N W_{ij}v^2_i\right). \qquad \label{rbm_amps_s1_new}
\end{eqnarray}
The inclusion of a quadratic visible bias now allows any product state in \eqr{eq:prod_state} to be described without hidden units by setting
\begin{equation*}
    a_j = \log\left(\sqrt{c^{(j)}_+c^{(j)}_-}/{c^{(j)}_0}\right) \quad {\rm and} \quad A_j = \log\left(\sqrt{c^{(j)}_+/c^{(j)}_-}\right),
\end{equation*}
while the quadratic interaction term ensures labelling freedom for the visible variable. 

\begin{figure}[ht]
\begin{center}
\includegraphics[scale=0.5]{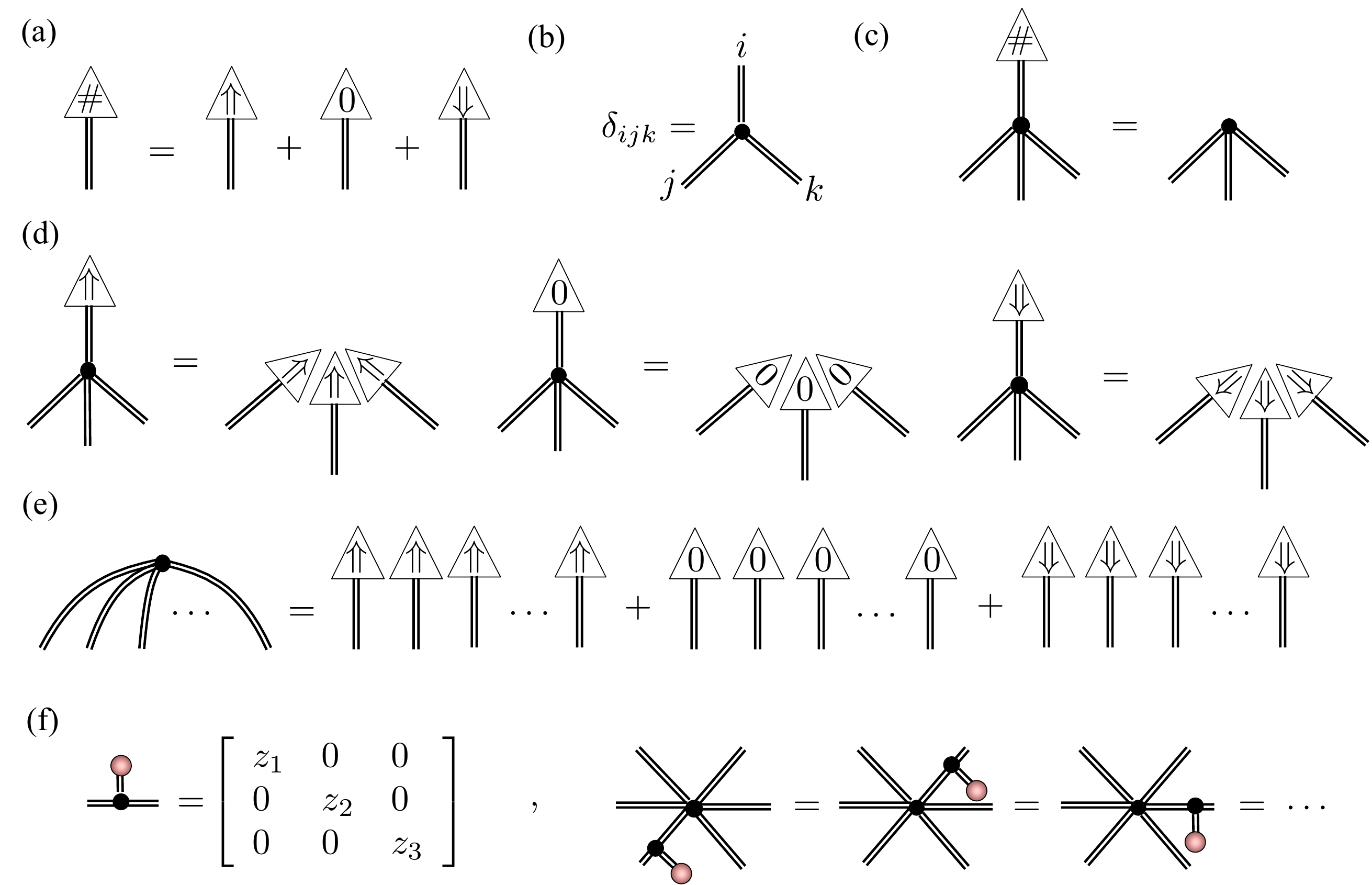}
\end{center}
\caption{A visual summary of the properties of the three-dimensional COPY tensor directly generalizing those outlined in \secr{sec:tnt} for the two-dimensional version.}
\label{fig:copy3d}
\end{figure}

A strong justification for \eqr{rbm_energy_s1} being the appropriate spin-1 generalization of RBMs is its relation to an NQS tensor network for spin-1. To handle spin-1 systems we introduce a COPY tensor with three-dimensional legs copying the $S^z$ basis states $\{\ket{\Uparrow},\ket{0},\ket{\Downarrow}\}$. Its properties are summarised in \fir{fig:copy3d}(a)-(f), and are straightforward generalisations of the spin-$\half$ case given in \fir{fig:copy2d}(a)-(f) and discussed in \secr{sec:tnt}. The major difference is that we now distinguish three-dimensional legs with `$=$' lines instead of `$\--$'. As before the NQS tensor network follows from the conversion of the RBM graph in \fir{fig:rbm}, except that visible vertices are now replaced by three-dimensional COPY tensors giving
\begin{equation}
\includegraphics[scale=0.5,valign=c]{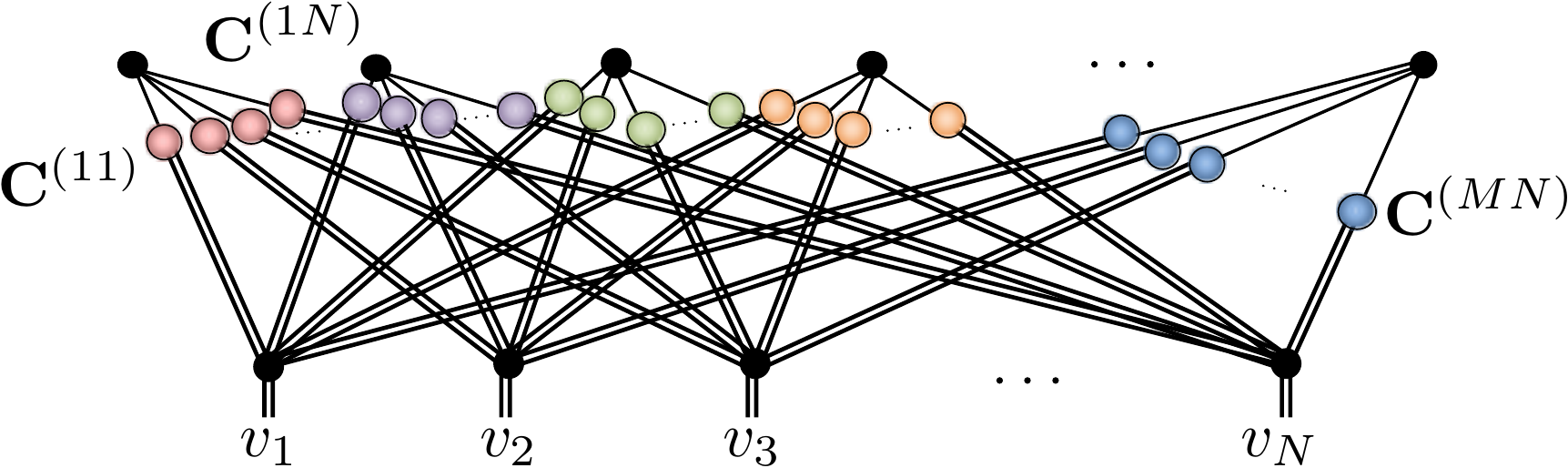} . \label{eq:nqs_spin1}
\end{equation}
Owing to the mixed dimensionality of the COPY tensors in this network it now requires a rectangular $2 \times 3$ coupling matrix ${\bf C}^{(ij)}$ between the $i$-th hidden and $j$-th visible unit. For the $i$-th hidden unit its set of coupling matrices $\Upsilon^{(i)} = \{{\bf C}^{(i1)},{\bf C}^{(i2)},\cdots,{\bf C}^{(iN)}\}$ can be explicitly tabulated as
\begin{equation*}
\begin{array}{rcccc}
& \overbrace{+1 \quad~~~ 0 \quad -1}^{\textstyle v_1} & \overbrace{+1 \quad~~~0\quad -1}^{\textstyle v_2} & \cdots & \overbrace{+1 \quad~~~0\quad -1}^{\textstyle v_N}  \\
\\
h_i\left\{\begin{array}{c}
+1 \\
-1 
\end{array}\right. &
\left[
\begin{array}{ccc}
{\scriptstyle C^{(i1)}_{++}} & {\scriptstyle C^{(i1)}_{+0}} & {\scriptstyle C^{(i1)}_{+-}}   \\
{\scriptstyle C^{(i1)}_{-+}} & {\scriptstyle C^{(i1)}_{-0}} & {\scriptstyle C^{(i1)}_{--}}
\end{array}
\right] & \left[
\begin{array}{ccc}
{\scriptstyle C^{(i2)}_{++}} & {\scriptstyle C^{(i2)}_{+0}} & {\scriptstyle C^{(i2)}_{+-}}   \\
{\scriptstyle C^{(i2)}_{-+}} & {\scriptstyle C^{(i2)}_{-0}} & {\scriptstyle C^{(i2)}_{--}}
\end{array}
\right] & \cdots & \left[
\begin{array}{ccc}
{\scriptstyle C^{(iN)}_{++}} & {\scriptstyle C^{(iN)}_{+0}}  & {\scriptstyle C^{(iN)}_{+-}}   \\
{\scriptstyle C^{(iN)}_{-+}} & {\scriptstyle C^{(iN)}_{-0}} & {\scriptstyle C^{(iN)}_{--}}
\end{array}
\right]
\end{array}.
\end{equation*}
The amplitudes $\Upsilon^{(i)}({\bm v})$ of this hidden unit's correlator then follow by summing the product of coupling matrix elements selected by $\bm v$ along each row, giving
\begin{equation}
\qquad \Upsilon^{(i)}({\bm v}) = \sum_{h_i=\pm 1}^1 \prod_{j=1}^N C^{(ij)}_{h_jv_i} = C^{(i1)}_{+v_1}C^{(i2)}_{+v_2}\cdots C^{(iN)}_{+v_N} + C^{(i1)}_{-v_1}C^{(i2)}_{-v_2}\cdots C^{(iN)}_{-v_N}.
\end{equation}
The amplitudes of NQS tensor network are then the product of each of these hidden unit correlators 
\begin{eqnarray}
\Psi_{\rm NQS}({\bm S} \rightarrow {\bm v}) &=& \prod_{i=1}^M \Upsilon^{(i)}({\bm v}), \label{eq:nqs_form}
\end{eqnarray}
which, like an RBM, can be exactly and efficiently sampled.

\par
The spin-1 NQS tensor network appears to have $5MN$ complex variational parameters, however again gauge freedom allows the shuffling of diagonal matrices (of an appropriate dimension) through a COPY tensors reducing this. Specifically, its equivalence to the generalized spin-1 RBM proposed in \eqr{rbm_energy_s1} is made using the coupling matrix decomposition
\begin{eqnarray*}
\left[
\begin{array}{ccc}
C^{(ij)}_{++} & C^{(ij)}_{+0} & C^{(ij)}_{+-}  \\
C^{(ij)}_{-+} & C^{(ij)}_{-0} & C^{(ij)}_{--}
\end{array}
\right] &=& e^c\left[
\begin{array}{cc}
e^{\tilde{b}_{ij}}  & 0  \\
0  & e^{-\tilde{b}_{ij}}   
\end{array}
\right] \left[
\begin{array}{ccc}
e^{w_{ij} + W_{ij}} & 1 & e^{-w_{ij} + W_{ij}}  \\
e^{-w_{ij} - W_{ij}} & 1 & e^{w_{ij} - W_{ij}}  
\end{array}
\right]\\
&&\qquad\qquad\qquad\qquad\quad \times \left[
\begin{array}{ccc}
e^{\tilde{a}_{ij} + \tilde{A}_{ij}}  & 0 & 0 \\
0 & 1 & 0 \\
0  & 0 & e^{-\tilde{a}_{ij} + \tilde{A}_{ij}}   
\end{array}
\right].
\end{eqnarray*}
The solution for the weights $w_{ij}, W_{ij}$ and partial biases $\tilde{a}_{ij},\tilde{A}_{ij},\tilde{b}_{ij}$ is outlined in App.~\ref{app:boltzmann}, from which the full RBM biases are formed as $a_j = \sum_{i=1}^M \tilde{a}_{ij}$, $A_j = \sum_{i=1}^M \tilde{A}_{ij}$ and $b_i = \sum_{j=1}^N \tilde{b}_{ij}$. The spin-1 NQS tensor network thus reduces to $n_{\rm params} = 2MN + 2N + M$ complex parameters. 

This correspondence between the spin-1 RBM and the spin-1 NQS tensor network highlights an advantage over unary encoding. Specifically, coupling matrices provide an intuitive tool for engineering the correlations and structures that a given hidden unit imprints on the amplitudes of an NQS. A trivial case is when all the elements of the $j$-th coupling matrix are 1's, denoted generically as $\bf I$, equivalent to the hidden unit being disconnected from that visible unit. A more complex example with conditional correlations is a hidden unit with coupling matrices 
\begin{equation*}
\{{\bf I}, \cdots, {\bf I}, \overbrace{{\bf C}_{{\rm c}\Uparrow}}^{k}, {\bf C}_{-0\Downarrow}, \cdots, {\bf C}_{-0\Downarrow}\},
\end{equation*}
built from
\begin{equation*}
{\bf C}_{{\rm c}\Uparrow} = \left[
\begin{array}{ccc}
0 & 1 & 1   \\
1 & 0 & 0
\end{array}
\right] \quad {\rm and} \quad {\bf C}_{-0\Downarrow} = \left[
\begin{array}{crr}
1 & 1 & 1  \\
1 & -1 & -1
\end{array}
\right].
\end{equation*}  
This hidden unit generates a correlator $\Upsilon({\bm v}) = \delta_{v_k,\Downarrow} + \delta_{v_k,\Uparrow}(-1)^{\mathcal{C}({\bm v})}$, where a factor $(-1)^{\mathcal{C}({\bm v})}$ is introduced conditional on the $k$th spin being in the state $\ket{\Uparrow}$, with $\mathcal{C}({\bm v})$ being the number of $\ket{0}$ and $\ket{\Downarrow}$ states in the configuration $\bm v$ between spins $k+1$ and $N$. Similar types of hidden units will be used extensively in \secr{sec:aklt_xyz} to construct an exact representation of a state.
\par

\subsection{Projection of unary encoding into a spin-1 RBM}\label{sec:unary2spin1}
The tensor network formalism provides further evidence that unary encoding from \secr{sec:rbm_unary} is an over-parameterisation of RBMs for spin-1 systems with $\delta n_{\rm params} = N(1+M)$ redundant parameters. Unary projection is implemented by an order-4 tensor $\mathcal{U}$ obeying
\begin{equation}
\includegraphics[scale=0.45,valign=c]{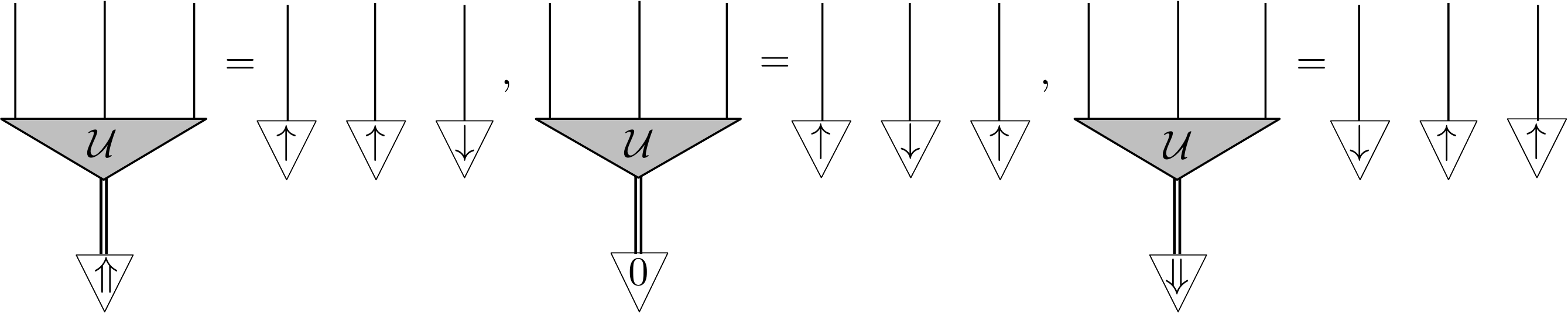} .
\end{equation}
By directly applying this projector to the spin-$\half$ NQS tensor network for a unary encoded state and performing a graphical rewrite we obtain the spin-1 NQS introduced in \eqr{eq:nqs_spin1}. \fir{fig:proj_unary} summarises the crucial manipulations required. 
\par

\begin{figure}[ht]
\begin{center}
\includegraphics[scale=0.5]{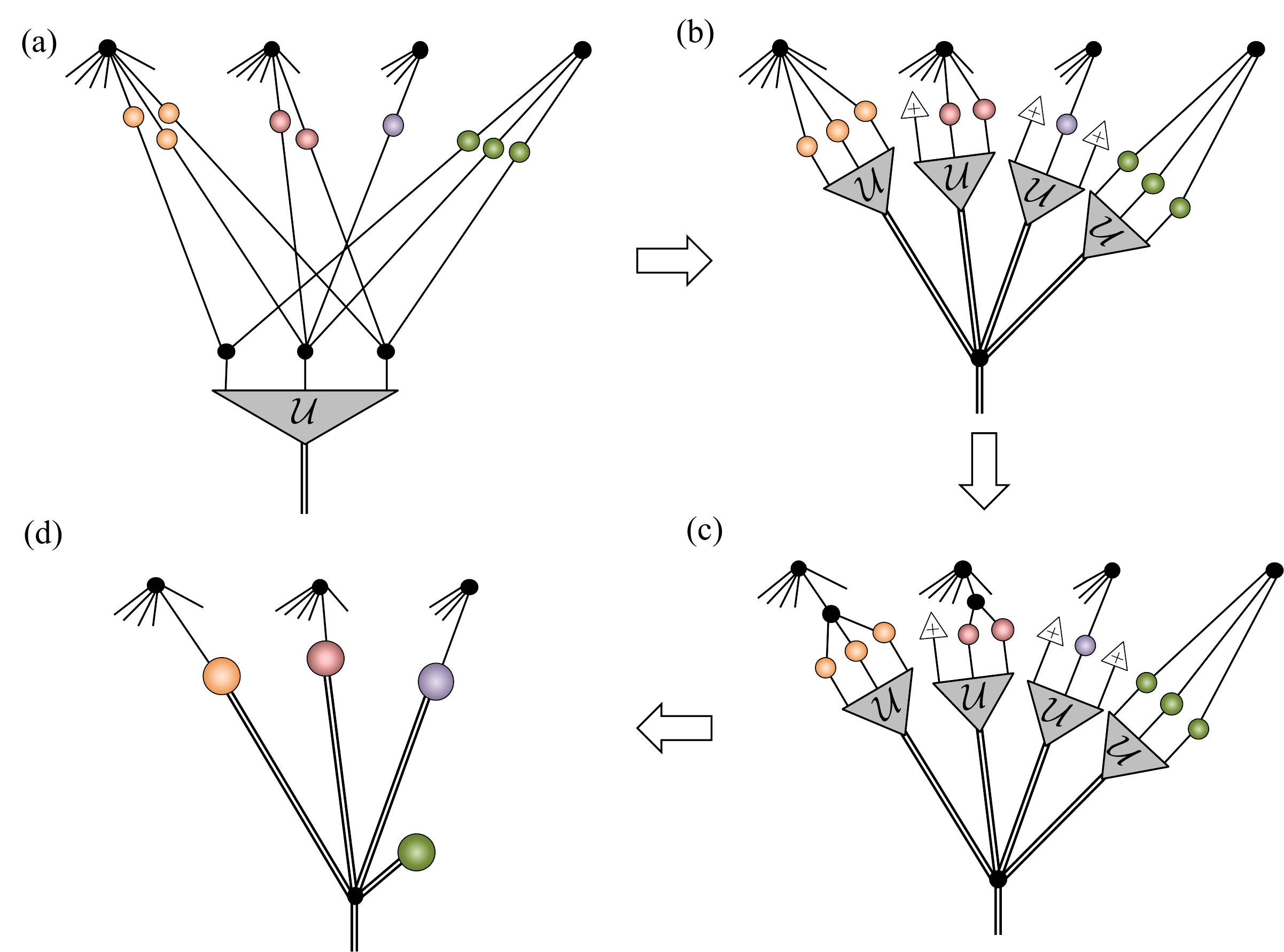}
\end{center}
\caption{(\textbf{a}) Representative examples of unary projection contractions with hidden units possessing different patterns of connectivity. (\textbf{b}) The first step in the contraction involves pulling the projection through the two-dimensional COPY tensors, leaving behind a single three-dimensional COPY tensor. (\textbf{c}) Connections to the unary spins are isolated by splitting the hidden unit COPY tensors. (\textbf{d}) The resulting blocks of the network are then contracted to obtain the $2 \times 3$ spin-1 NQS coupling matrices.}
\label{fig:proj_unary}
\end{figure}

In \fir{fig:proj_unary}(a) some representative examples of contractions between the projection tensor $\mathcal{U}$ and hidden units are shown. There are three steps to rewriting the network. The first step, shown in \fir{fig:proj_unary}(b), essentially pulls $\mathcal{U}$ through the three unary two-dimensional COPY tensors, leaving behind a single three-dimensional COPY tensor representing the physical spin-1. Two important cases are shown in the example in \fir{fig:proj_unary}(b). A hidden unit may have connections to each of the unary spin-$\half$'s, where upon they get bundled up by the $\mathcal{U}$ tensor. A hidden unit may connect to only a subset of the unary spin-$\half$'s, which is handled by plugging the unused legs of $\mathcal{U}$ tensor with $\ket{+}$. If a hidden unit couples to more than one of the unary spins then the second step, shown in \fir{fig:proj_unary}(c), involves splitting the hidden unit's two-dimensional COPY tensor to separate those connections. The final step is then to contract the split COPY tensor, coupling matrices and the projection $\mathcal{U}$ to form a rectangular $2 \times 3$ coupling matrix, as depicted in \fir{fig:proj_unary}(d). If a hidden unit has connections exclusively within the unary spin-$\half$'s then it becomes an entirely local visible bias contribution in the spin-1 NQS. 
\par

\subsection{Change of local spin basis}
For tensor network representations, such as MPS or PEPS, the complexity (internal bond dimension) of a given state's description is rooted in its entanglement structure. As such changing the local basis of the spins used in a calculation has no effect on this complexity. Moreover, transforming a representation from one basis to another is accomplished by simply admixing the local tensors. Within VMC a change of local spin basis leaves the locality and the sparsity of the Hamiltonian essentially unchanged. However, it is pivotal to the method that the amplitudes $\Psi_{\bm p}({\bm S})$ of whatever ansatz is used can be efficiently evaluated in this new basis. This is not generally true of NQS since their sampleability is intimately tied to the basis that factorises the COPY tensors they are built from.
\par
To understand how NQS behave consider a representation of some state $\ket{\Psi}$, for instance of four spin-1's sampleable in the $S^z$ basis
\begin{equation}
\includegraphics[scale=0.45,valign=c]{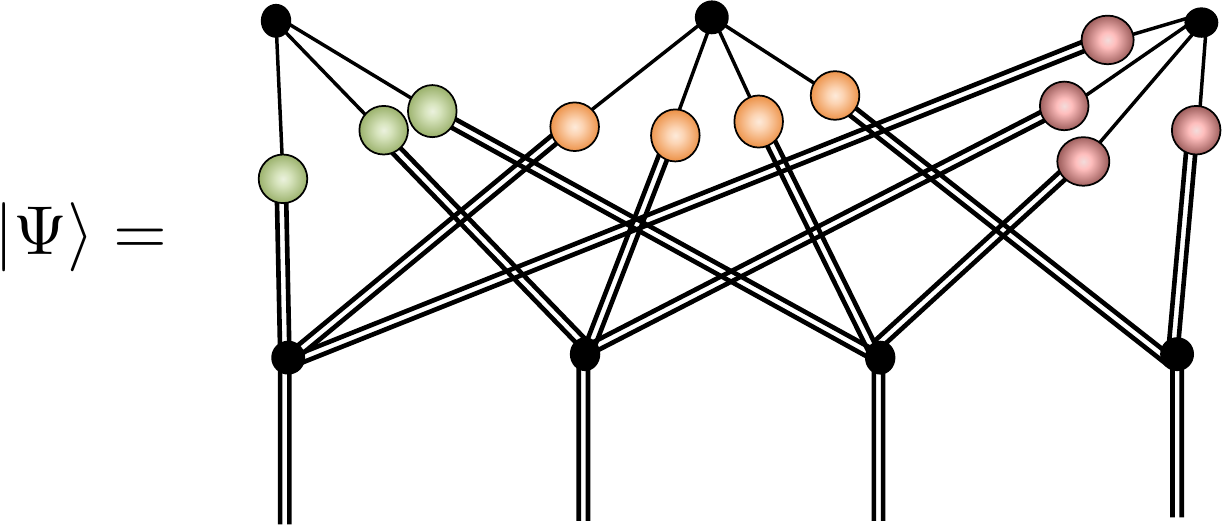} .
\end{equation}
Now suppose we transform the local $S^z$ basis $\ket{S}$ to a new basis $\chi$ via a unitary $\ket{\chi_S} = \hat{\mathcal{B}}^\dagger\ket{S}$. Formally, from a tensor network perspective we find the $\chi$ basis NQS representation by sandwiching $\mathbbm{1} = \hat{\mathcal{B}}\hat{\mathcal{B}}^\dagger$ on each physical leg and computing the $S^z$ basis NQS representation of $(\hat{\mathcal{B}}^{\dagger})^{\otimes N}\ket{\Psi}$. However, currently there is no known procedure for updating exactly an NQS after the application of an arbitrary single-spin unitary, even allowing for an increased number of hidden units. While we are guaranteed that an NQS representation of $\hat{\mathcal{B}}^{\otimes N}\ket{\Psi}$ exists, as illustrated here schematically
\begin{equation}
\includegraphics[scale=0.45,valign=c]{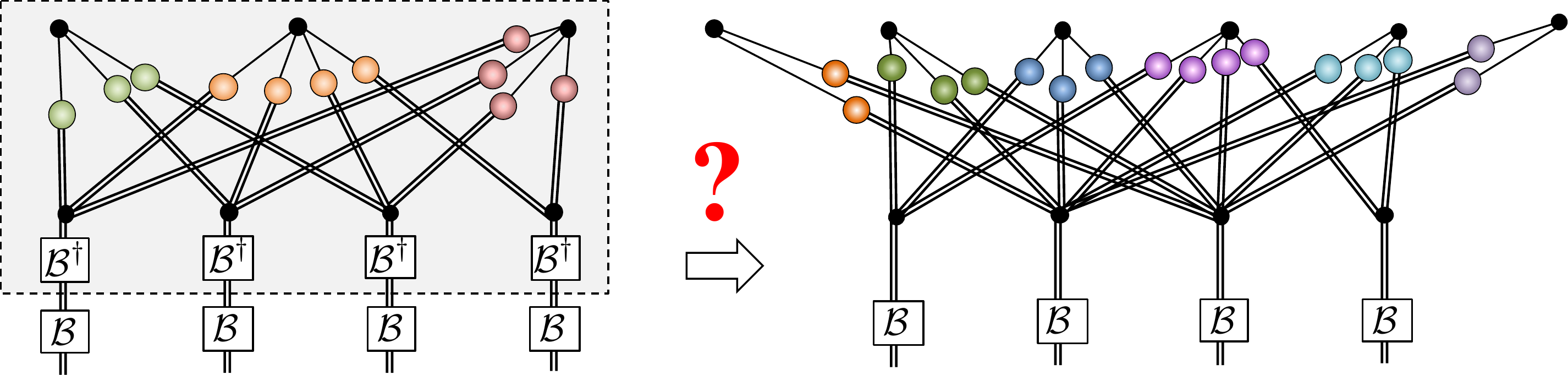} ,
\end{equation}
there is no guarantee it will be efficient, even if the representation of $\ket{\Psi}$ was originally. An example of such a catastrophic loss of NQS efficiency has been presented by Gao and Duan~\cite{gao_dbm17}. They show that, so long as the polynomial hierarchy in computational complexity theory does not collapse, a two-dimensional cluster state, which has an efficient NQS representation, has no efficient NQS representation after a specific layer of translation-invariant single-spin unitaries are applied. Thus NQS complexity depends non-trivially on the local spin basis used.
\par
Motivated by this we will consider NQS calculations in two different local spin-1 bases to examine how the complexity varies. Specifically, the standard $S^{z}$ basis and the $xyz$ basis defined as
\begin{eqnarray*}
\ket{x} = \frac{1}{\sqrt{2}}\left(\ket{\Uparrow} - \ket{\Downarrow}\right), \quad \ket{y} = \frac{\rm i}{\sqrt{2}}\left(\ket{\Uparrow} + \ket{\Downarrow}\right), \quad \ket{z} = -\ket{0}.
\end{eqnarray*}
In the $xyz$ basis the individual spin-1 operators all acquire the same off-diagonal form
\begin{eqnarray*}
\quad \hat{S}^x_j &=& \left[
\begin{array}{ccc}
0  & 0  & 0  \\
0  & 0  & {\rm i} \\
0  & -{\rm i}  & 0  
\end{array}
\right], \quad 
\hat{S}^y_j = \left[
\begin{array}{ccc}
0  & 0  & {\rm i}  \\
0  &  0 & 0  \\
-{\rm i}  & 0 & 0  
\end{array}
\right], \quad
\hat{S}^z_j = \left[
\begin{array}{ccc}
0  & {\rm i}  & 0  \\
-{\rm i}  & 0  & 0  \\
0  & 0  & 0  
\end{array}
\right],
\end{eqnarray*}
as a consequence of them all contributing one eigenstate to the basis. On a practical level using the $xyz$ basis for the spin-1 RBM in \eqr{rbm_amps_s1_new} simply requires replacing $\bm S$ with $\bm\alpha$ and a physical-visible mapping from ${\bm \alpha} \mapsto {\bf v}$ such as
\begin{equation}
\alpha_j = \left\{
\begin{array}{ccr}
x  & \mapsto & +1    \\
y  & \mapsto & 0  \\
z & \mapsto & -1    
\end{array}\right\} = v_j.
\end{equation}
Equivalently, the visible unit COPY tensors for this NQS tensor network can be considered to be rotated into this $xyz$ basis.
\par

\subsection{Numerical example - Spin-1 anti-ferromagnetic Heisenberg model} \label{sec:afh}
To confirm the effectiveness of our spin-1 NQS we performed VMC calculations to reach the ground state of the well-known anti-ferromagnetic Heisenberg (AFH) model in one dimension. The Hamiltonian is given by
\begin{equation}
    \hat{\mathcal{H}} = J\sum_{i} \hat{\bf S}_{i}\cdot \hat{\bf S}_{i+1},
\end{equation}
where $J > 0$ is the magnetic interaction strength and with periodic boundary conditions $N+1 \equiv 1$. We focus on small systems allowing direct comparison to the ground state calculated from exact diagonalisation via the overlap $\mathcal{O} = |\braket{\Psi_{\rm ED}}{\Psi_{\bm \Lambda}}|^2$. Additionally, we performed NQS optimisations in both the $S^{z}$ and $xyz$ bases to compare any differences in performance. This basis change alters the fixed quantum numbers of the system. In particular, in the $S^{z}$ basis the AFH model preserves the total $S^{z}$ projection of the system and its ground state lies in the $\sum_{j=1}^{N}S^{z}_{j} = 0$ sector of the full Hilbert space.  While in the $xyz$ basis the AFH model preserves the ``parity" of the total $x, y, z$ spin populations in the system and for even $N$ the ground state lies in the subspace of configurations where there is an even number of each basis state. As is common for VMC calculations we only select configurations from the relevant subspace during sampling. We perform the calculations with hidden unit numbers $M = [1, 2, \dots , 2N]$, initialising $M = 1$ with random small complex parameters. For each successive calculation we seed the NQS with the parameters for $M-1$ and initialise the $M$th hidden unit with random small parameters, gradually increasing the size of the network in a sequential manner. To check the robustness against initialisation bias of the qualitative features we discuss we rerun optimisation sequences 5 to 10 times and present the best results here.
\par
\begin{figure}[ht]
\begin{center}
\includegraphics[scale=0.5]{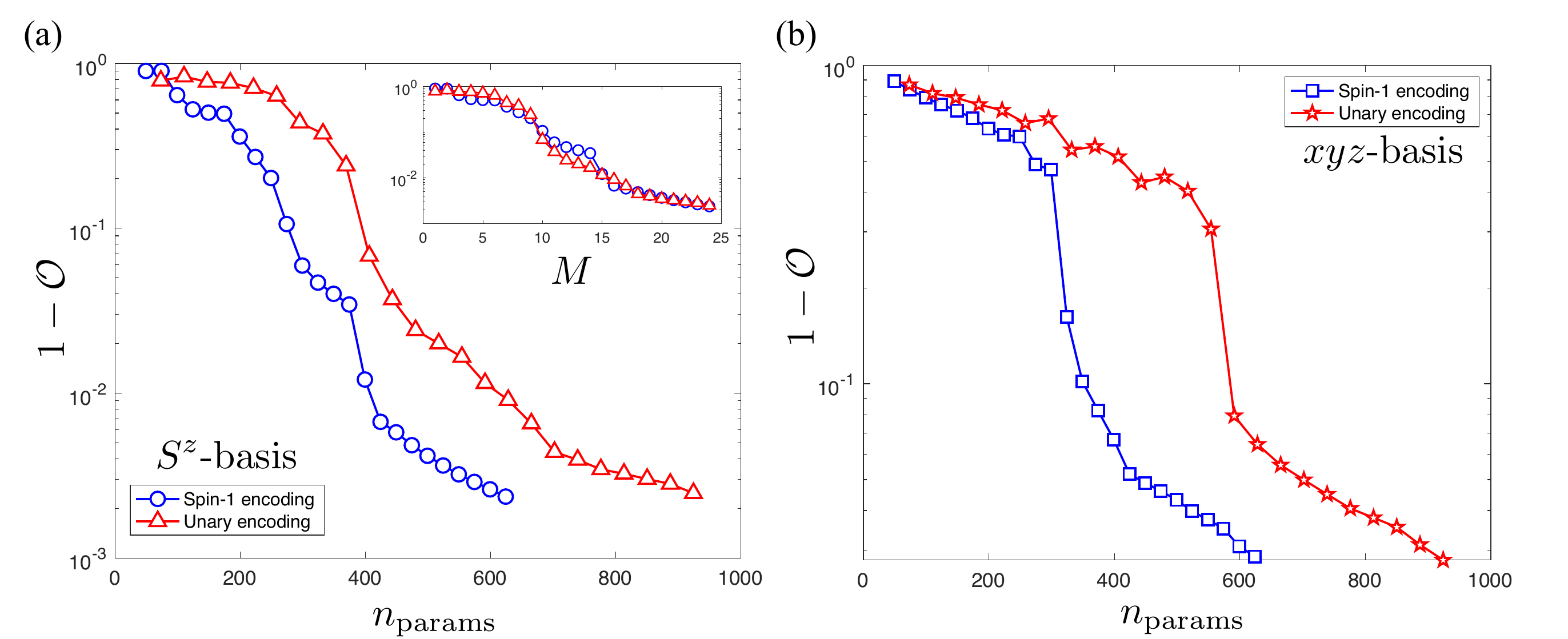}
\end{center}
\caption{Plots of the infidelity of unary (red) and spin-1 NQS (blue) with the spin-1 AFH ground state with periodic boundary conditions. The calculations presented in both plots have the same number of sites $N = 12$, up to $M = 2N = 24$ hidden units. (\textbf{a}) The infidelity $1 - \mathcal{O}$ of the two NQS formulations for the $S^z$ basis versus the RBM parameter count $n_{\rm params}$. The inset shows the collapse of the same data when it is plotted in terms of the hidden unit number $M$. (\textbf{b}) The same calculations performed in the $xyz$ basis.}
\label{fig:afh_numeric}
\end{figure}
In \fir{fig:afh_numeric} we show how the accuracy of spin-1 NQS and unary encoding representations improve with an increasing number of hidden units $M$ for both the $S^{z}$ and $xyz$ bases plotted in terms of the variational parameter count. The spin-1 NQS achieves a superior accuracy to unary encoding for a similar $n_{\rm params}$. The inset of \fir{fig:afh_numeric}(a) shows the collapse of the same $S^z$ data plotted against $M$, indicating that the spin-1 NQS and unary encoding have in fact located the same solution for a given $M$. However by using $\delta n_{\rm params} = N(1+M)$ less parameters the spin-1 NQS is considerably more efficient to optimise, especially noting that $\delta n_{\rm params}$ scales with both system size and hidden unit number\footnote{For example consider that the $N = 12, M = 12$ spin-1 NQS has a comparable parameter count to an $N = 12, M = 8$ unary NQS.}. In \fir{fig:afh_numeric}(b) we observe a noticeable drop in accuracy for both NQS variants in the $xyz$ basis compared to the $S^{z}$ basis. This suggests that the AFH ground state amplitude structure with periodic boundaries is inherently more complicated in the $xyz$ basis regardless of encoding. Moreover, this confirms that hidden unit number $M$ of an NQS is basis-dependent quantity and cannot be used as a proxy of the entanglement.
\par

\section{Revisiting the AKLT model} \label{sec:aklt}
We now move on to benchmark our spin-1 NQS against the analytically solvable AKLT model~\cite{affleck87} which is a spin-1 chain governed by a bilinear-biquadratic SU(2)-isotropic Heisenberg Hamiltonian of the form
\begin{equation*}
\hat{H}_{\rm AKLT} = \sum_{j=1}^N\left[\hat{\bf S}_j\cdot \hat{\bf S}_{j+1} + \beta\left(\hat{\bf S}_j\cdot \hat{\bf S}_{j+1}\right)^2\right] + \frac{2}{3},
\end{equation*}
with periodic boundary conditions $N+1 \equiv 1$. It has special significance since it was the first solvable spin-1 chain model that exhibits the `Haldane gap'~\cite{haldane83}. The AKLT state $\ket{\Psi_{\rm AKLT}}$ is the ground state of $\hat{H}_{\rm AKLT}$ at the AKLT point $\beta = \frac{1}{3}$, and has an energy of exactly zero. 

\begin{figure}[ht]
\begin{center}
\includegraphics[scale=0.35]{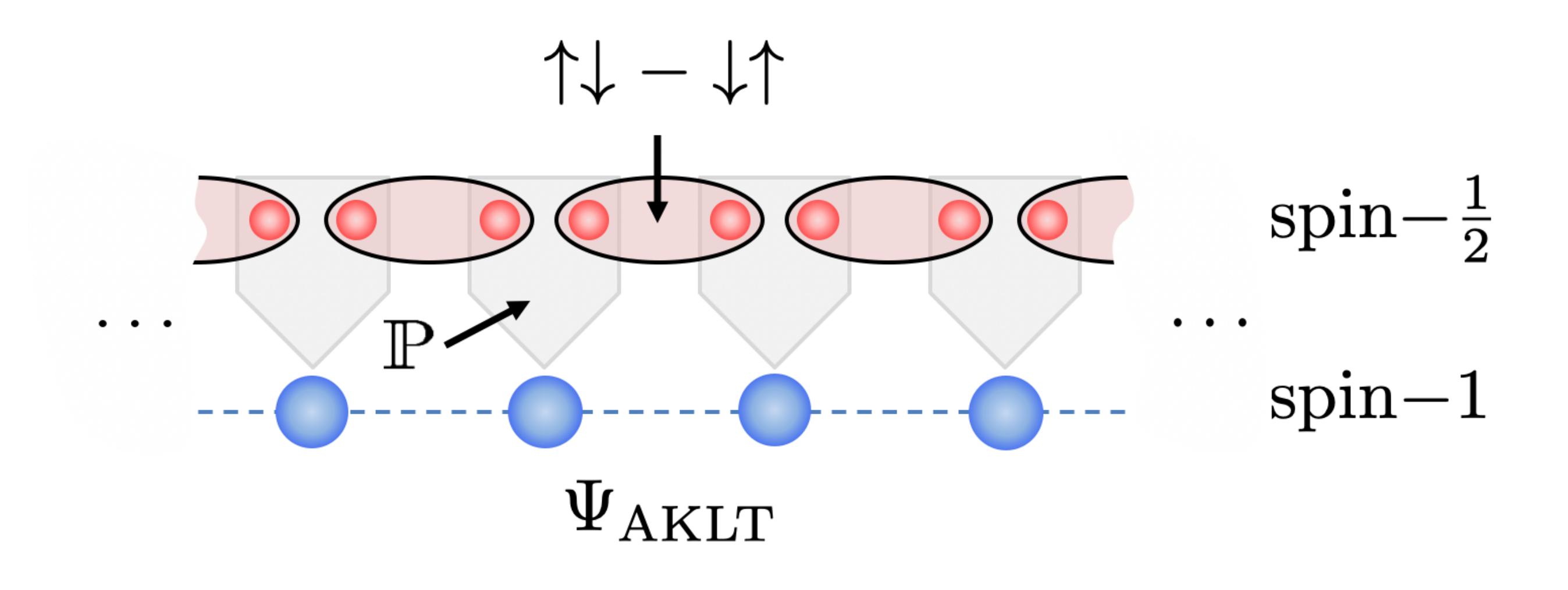}
\end{center}
\caption{Valence bond solid construction of the spin-1 AKLT state $\Psi_{\rm AKLT}$ from the projection $\mathbbm{P}$ of pairs of spin-$\half$ particles shared between neighbouring sites on the chain.}
\label{fig:aklt}
\end{figure}

As is well-known, $\ket{\Psi_{\rm AKLT}}$ has a special structure of correlations which are related to a valence bond solid. Specifically, each spin-1 is envisaged as being a pair of spin-$\half$ particles that are correspondingly entangled in a singlet state with a partner spin-$\half$ in the nearest neighbouring spin-1 on the chain. The AKLT state is then the projection $\mathbbm{P}$ of the local pairs of spin-$\half$ particles into the triplet subspace, as depicted in \fir{fig:aklt}. This also leads to $\ket{\Psi_{\rm AKLT}}$ possessing a very compact MPS representation with matrices
\begin{eqnarray*}
{\bf A}^{+} &=& \frac{1}{\sqrt{2}}\hat{\sigma}^+, \quad {\bf A}^{0} = -\frac{1}{2}\hat{\sigma}^z, \quad {\bf A}^{-} = -\frac{1}{\sqrt{2}}\hat{\sigma}^-,
\end{eqnarray*}
where $\hat{\sigma}^\pm = \half(\hat{\sigma}^x \pm {\rm i}\hat{\sigma}^y)$, such that the (unnormalized) amplitudes of the ground state in the $\hat{S}^z$ basis follow as
\begin{equation}
\Psi_{\rm AKLT}({\bm S}) = {\rm tr}\left({\bf A}^{S_1}{\bf A}^{S_2}\cdots {\bf A}^{S_N}\right). \label{eq:aklt_sz_amps}
\end{equation}
Since the AKLT point of $\hat{H}_{\rm AKLT}$ lies in the gapped Haldane phase $\ket{\Psi_{\rm AKLT}}$ has finite-ranged magnetic correlations,
\begin{equation}
O^{zz}_\ell = \bra{\psi_{\rm AKLT}}\hat{S}^z_0\hat{S}^z_\ell\ket{\psi_{\rm AKLT}} \sim e^{-\ell/\xi}, \quad {\rm with} \quad \xi = \frac{1}{{\rm ln}(3)},
\end{equation}
yet it also has an unbroken spin rotation symmetry which is a hallmark of a symmetry protected topological order. Specifically, the string-order parameter
\begin{equation}
O^{\rm string} = \lim_{N\rightarrow\infty} \lim_{\ell \rightarrow\infty}\bra{\psi_{\rm AKLT}}\hat{S}^z_0\prod_{j=1}^{\ell-1}e^{i\pi\hat{S}^z_j}\hat{S}^z_\ell\ket{\psi_{\rm AKLT}} \sim -\frac{4}{9},
\end{equation}
reveals the presence of infinite-ranged antiferromagnetic correlations. This is evident from the structure of the MPS amplitudes. Any matrix product ${\bf A}^{\pm}{\bf A}^0 \cdots {\bf A}^0{\bf A}^{\pm} = 0$, so any configuration containing a ferromagnetic segment, like “+ 0 0 0 +”, with any number of 0's is not allowed. In contrast allowed configurations contain only antiferromagnetic segments, such as “\textendash\,0 + 0 0 0 \textendash\ + 0”, arising from sequences like ${\bf A}^{\pm}{\bf A}^0 \cdots {\bf A}^0{\bf A}^{\mp}$ where every $\pm$ is partnered with a $\mp$ separated by with an arbitrary string of 0's.  
\par
Despite its simple MPS representation it is surprisingly non-trivial to capture the non-commutative matrix products making up the AKLT amplitudes with an NQS. Direct conversion to an NQS from the MPS representation gives two reasons why it must contain long-ranged hidden units. First, it has been shown~\cite{zheng_stbl19} that any short-ranged translationally invariant NQS cast into a MPS form by mapping hidden units into virtual bonds has $\bf A$ matrices that are at most rank-1. Since the matrix ${\bf A}^0$ in the AKLT state is rank-2 it fails this condition. Second, if we divide the chain into a sequence of three contiguous parts $a,b,c$, once we make $b$ larger than the longest range of any hidden unit, so no hidden unit connects to visible units in both $a$ and $c$, then the NQS amplitudes factorise as
\begin{equation}
\psi_{\rm sr-NQS}({\bm S}_a,{\bm S}_b,{\bm S}_c) =  \psi_{ab}({\bm S}_a,{\bm S}_b) \psi_{bc}({\bm S}_b,{\bm S}_c),
\end{equation}
implying that ${\bm S}_a$ and ${\bm S}_c$ are uncorrelated~\cite{chen18}. The AKLT amplitudes $\Psi_{\rm AKLT}({\bm S})$ do not satisfy this property since region $b$ can be any length of $0$'s and there will always be non-zero amplitudes “+ 0 0 0 \dots 0 \textendash” and “\textendash\, 0 0 0 \dots 0 +” encoding string order correlations that are not factorisable. 
\par
It has been previously found that the AKLT state in the $S^z$ basis requires an NQS with $M \sim O(N^2)$ long-ranged hidden units~\cite{glasser18}, and this was borne out in numerical calculations for small systems. In App.~\ref{app:aklt_sz} we explicitly construct an NQS for the $S^z$ basis AKLT amplitudes using $M = 2N^2 + N \lfloor\half(N-1)\rfloor + 1$ hidden units, many of which are extensive over the system. The $O(N^2)$ scaling can be readily understood as a consequence of having hidden units that each eliminate disallowed configurations, such as ``$\pm$ 0 0 0 $\pm$", and impose the sign for allowed configurations, such as ``$\pm$ 0 0 0 $\mp$", for all $N$ separations and $N$ translations over the system. This is rather less efficient than the compact spin-$\half$ NQS found for Jastrow, graph and stabilizer states in Ref.~\cite{Pei21}. The AKLT state can be expressed with $O(N)$ hidden units but at the expense of needing a 2-layer DBM network~\cite{zheng_stbl19} that cannot in general be exactly sampled, complicating its use numerically. However, as we saw for the AFH model numerical results the hidden unit complexity is basis dependent. Surprisingly, we will show next that an exact $M \sim O(N)$ spin-1 NQS representation of the AKLT state is obtained in the $xyz$ basis.
\par

\subsection{Exact spin-1 NQS for AKLT state in the $xyz$ basis} \label{sec:aklt_xyz}
The AKLT state provides an instructive example of how a single spin basis change can significantly alter the amplitude structure. Transforming the MPS representation into the $xyz$ basis yields matrices
\begin{eqnarray*}
~{\bf B}^{x} &=& \frac{1}{\sqrt{2}}({\bf A}^{+} - {\bf A}^{-1}) = \frac{1}{2}\hat{\sigma}^x, \quad {\bf B}^{y} = \frac{-{\rm i}}{\sqrt{2}}({\bf A}^{+1} + {\bf A}^{-1}) = \frac{1}{2}\hat{\sigma}^y, \quad {\bf B}^{z} = -{\bf A}^0 = \frac{1}{2}\hat{\sigma}^z,
\end{eqnarray*}
and thus renders the amplitudes into products of Pauli matrices
\begin{equation*}
\Psi_{\rm AKLT}({\bm \alpha}) = {\rm tr}\left({\bf B}^{\alpha_1}{\bf B}^{\alpha_2}\cdots {\bf B}^{\alpha_N}\right),
\end{equation*}
where ${\bm \alpha} = (\alpha_1,\dots,\alpha_N)$ with $\alpha_j = \{x,y,z\}$ label the $xyz$ basis. As expected there is no change in the complexity/internal dimension of the MPS representation.

\begin{figure}[ht]
\begin{center}
\includegraphics[scale=0.4]{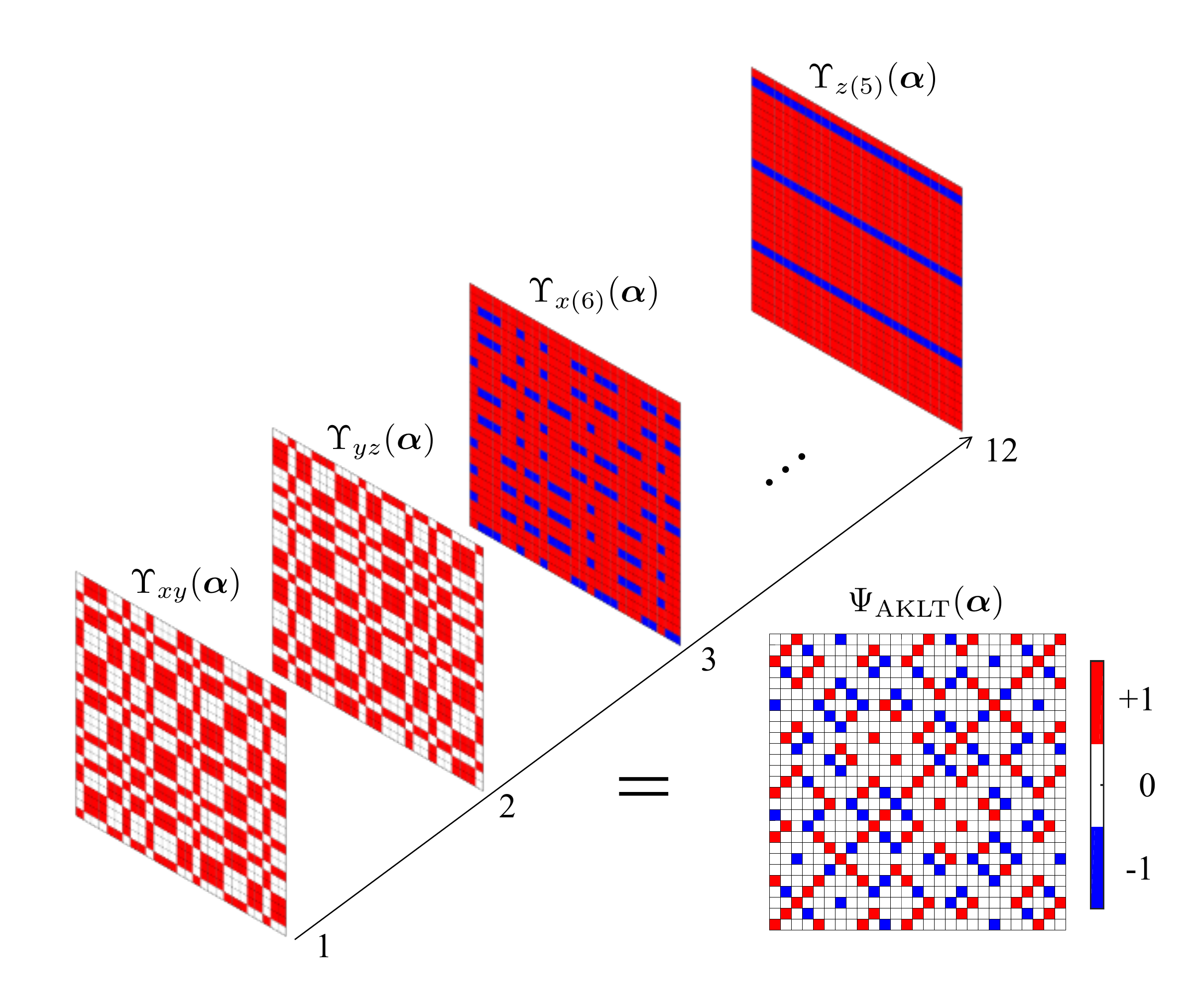}
\end{center}
\caption{The $3^6$ (unnormalised) amplitudes of the $N=6$ AKLT state in the $xyz$ basis $\Psi_{\rm AKLT}({\bm \alpha})$ are displayed here as a $3^3 \times 3^3$ matrix with the colour of an element designating which of the values $+1,0,-1$ the amplitude has. The bottom left corner element corresponds to the amplitude of $\ket{xxxxxx}$, while the top right corner corresponds to $\ket{zzzzzz}$. The amplitudes $\Psi_{\rm AKLT}({\bm \alpha})$ are reconstructed exactly by the product of $M=2N=12$ hidden unit filters given in the main text shown. The first two filters $\Upsilon_{xy}({\bm \alpha})$ and $\Upsilon_{yz}({\bm \alpha})$ establish the nodal structure, while the other ten filters $\Upsilon_{x(k)}({\bm \alpha})$ and $\Upsilon_{z(k)}({\bm \alpha})$ imprint the sign structure.}
\label{fig:aklt_xyz_amps}
\end{figure}

The structure of the amplitudes $\Psi_{\rm AKLT}({\bm \alpha})$ in the $xyz$ basis is significantly simpler than $\Psi_{\rm AKLT}({\bm S})$ in the $S^z$ basis. Amplitudes are now evaluated by tracking the anticommutations of Pauli matrices required to make the matrices of each the type form a contiguous sequence, e.g. $xx\cdots xyy\cdots yzz\cdots z$, and then reducing the product repeatedly via $(\hat{\sigma}^\alpha)^2 = \mathbbm{1}$. The resulting matrix trace is non-zero only when the overall product is $\propto \mathbbm{1}$, and so all non-zero amplitudes have an equal magnitude. Depending on whether $N$ is even or odd, this condition requires that there is either an even or odd number of $x,y$ and $z$'s in any configuration string, respectively. Using this we arrive at the following result:

\begin{Theorem}[AKLT state $xyz$ NQS] \label{thm:aklt_xyz}
The AKLT state in the $xyz$ spin basis has an exact spin-1 NQS representation requiring $M=2N$ hidden units.
\end{Theorem}

\begin{proof}
We establish this result using a direct and intuitive construction for $\Psi_{\rm AKLT}({\bm \alpha})$ in which hidden units are devised to implement the nodal structure and sign structure of this state. The rules governing the amplitudes are as follows:
\begin{enumerate}
\item To implement the parity constraint on the number of $x,y$ and $z$'s in any configuration string $\bm \alpha$ we introduce the following $2 \times 3$ coupling matrices
\begin{equation*}
{\bf C}_{xy} = \left[
\begin{array}{rrc}
1 & 1 & 1  \\
-1 & -1 & 1   
\end{array}
\right], \quad {\bf C}_{yz} = \left[
\begin{array}{crr}
1 & 1 & 1  \\
1 & -1 & -1
\end{array}
\right].
\end{equation*}
By defining two hidden units from these matrices as $\Upsilon_{xy} = \{{\bf C}_{xy}, {\bf C}_{xy}, \cdots, {\bf C}_{xy}\}$ and $\Upsilon_{yz} = \{{\bf C}_{yz}, {\bf C}_{yz}, \cdots, {\bf C}_{yz}\}$ we arrive at the product filter
\begin{equation*}
\Upsilon_{xy}({\bm \alpha})\Upsilon_{yz}({\bm \alpha}) = \left(1 + (-1)^{\#\,x{\rm 's}~+~\#\,y{\rm 's}}\right)\left(1+ (-1)^{\#\,y{\rm 's}~+~\#\,z{\rm 's}}\right),
\end{equation*}
in which the hidden units cancel out any strings $\bm \alpha$ that have odd numbers of both $x$'s and $y$'s, and $y$'s and $z$'s, respectively. Together these hidden units completely establish for any $N$ the nodal structure of the AKLT state amplitudes in this basis.
\item To reproduce the sign structure arising from anticommuting Pauli matrices into a contiguous sequence we require two types of hidden units. The first type of hidden unit uses a conditional coupling matrix for the local state $\ket{x}$
\begin{equation*}
{\bf C}_{x} = \left[
\begin{array}{crc}
0 & 1 & 1  \\
1 & 0 & 0   
\end{array}
\right],
\end{equation*}
along with ${\bf C}_{yz}$ to define a hidden unit of the form 
\begin{equation*}
\Upsilon_{x(k)} = \{{\bf C}_{yz}, \cdots, {\bf C}_{yz}, \overbrace{{\bf C}_x}^{k},{\bf I}, \cdots, {\bf I}\},
\end{equation*}
where ${\bf C}_x$ appears in the $k$th position in the sequence. The action of $\Upsilon_{x(k)}$ is to induce on a configuration a factor $(-1)^{\#\,y{\rm 's}~+~\#\,z{\rm 's}}$ between site $k-1$ and the left boundary, conditional on site $k$ being in state $\ket{x}$. This is the sign that would occur if a $\sigma^x$ matrix was anticommuted to this boundary through the corresponding product of Pauli matrices. The second type of hidden unit uses two further coupling matrices 
\begin{equation*}
{\bf C}_{y} = \left[
\begin{array}{crc}
1 & 1 & 1  \\
1 & -1 & 1   
\end{array}
\right], \quad {\bf C}_{z} = \left[
\begin{array}{ccc}
1 & 1 & 0  \\
0 & 0 & 1
\end{array}
\right],
\end{equation*}
defining a hidden unit of the form
\begin{equation*}
\Upsilon_{z(k)} = \{{\bf I}, \cdots, {\bf I}, \overbrace{{\bf C}_{z}}^{k}, {\bf C}_{y}, \cdots, {\bf C}_{y}\},
\end{equation*}
where ${\bf C}_{z}$ appears in the $k$th position in the sequence. The action of $\Upsilon_{z(k)}$ is to induce on a configuration a factor $(-1)^{\#\,y{\rm 's}}$ between site $k+1$ and the right boundary, conditional on site $k$ being in state $\ket{z}$. This is the sign that would occur if a $\hat{\sigma}^z$ matrix was anticommuted to the right boundary through the corresponding product of Pauli matrices, assuming that any $\hat{\sigma}^x$'s have already been anticommuted to the left boundary. To capture all locations $k$ for both types thus requires $2(N-1)$ hidden units which entirely establish the sign structure of the AKLT state amplitudes in this basis.
\end{enumerate}
This gives a total of $M=2N$ hidden units.\end{proof}
\noindent The resulting amplitude-wise product decomposition of the AKLT into hidden unit correlators is depicted \fir{fig:aklt_xyz_amps} for $N=6$. 
\par

\subsection{Analytic example - AKLT unary stabilizer state} \label{sec:unary_state}
An explicit NQS construction for the AKLT state in the $xyz$ basis has been given before in Ref.~\cite{lu19} using unary encoded cells of $\{{\rm a, b, c}\}$ spin-$\half$'s. Their construction involves initialising the b spin-$\half$'s in state $\ket{+}$ while entangling the a and c spin-$\half$'s between adjacent unary cells in the state $\ket{\uparrow}\ket{\uparrow} + \ket{\downarrow}\ket{\downarrow}$. As a tensor network this is represented as
\begin{equation}
\includegraphics[scale=0.45,valign=c]{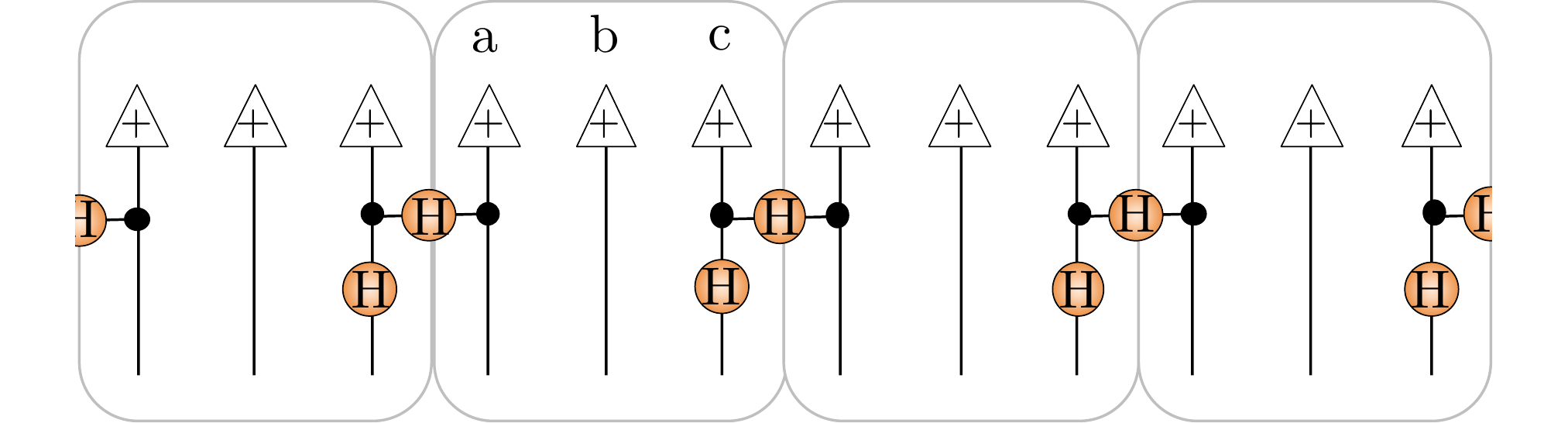} ,\label{eq:aklt_xyz_unary_init}
\end{equation}
where each box is a unary cell and H denotes the Hadamard matrix. Each unary cell then has the following unitary applied\footnote{We have switched the controlled-NOT gates in the circuit given in Figure 5(b) of Ref.\cite{lu19} into controlled-$Z$ gates here to expose the graph state equivalence.} to it 
\begin{equation}
\includegraphics[scale=0.45,valign=c]{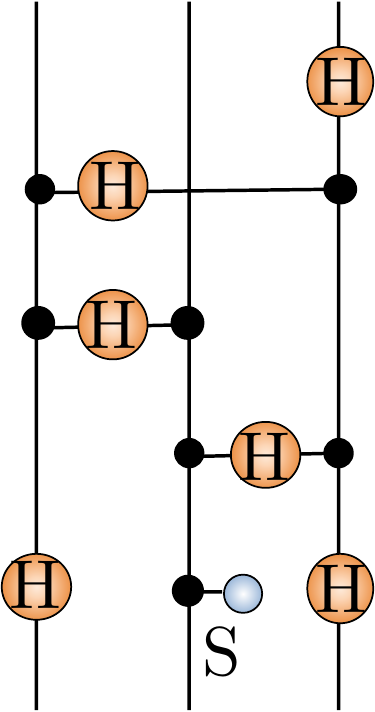} ,\label{eq:aklt_xyz_unary_unitart}
\end{equation}
where S is the phase gate~\cite{lu19}. Putting these pieces together the unary encoded spin-$\half$ state $\ket{\Psi_{\rm unary}}$ is a stabilizer state constructed by the circuit
\begin{equation}
\includegraphics[scale=0.45,valign=c]{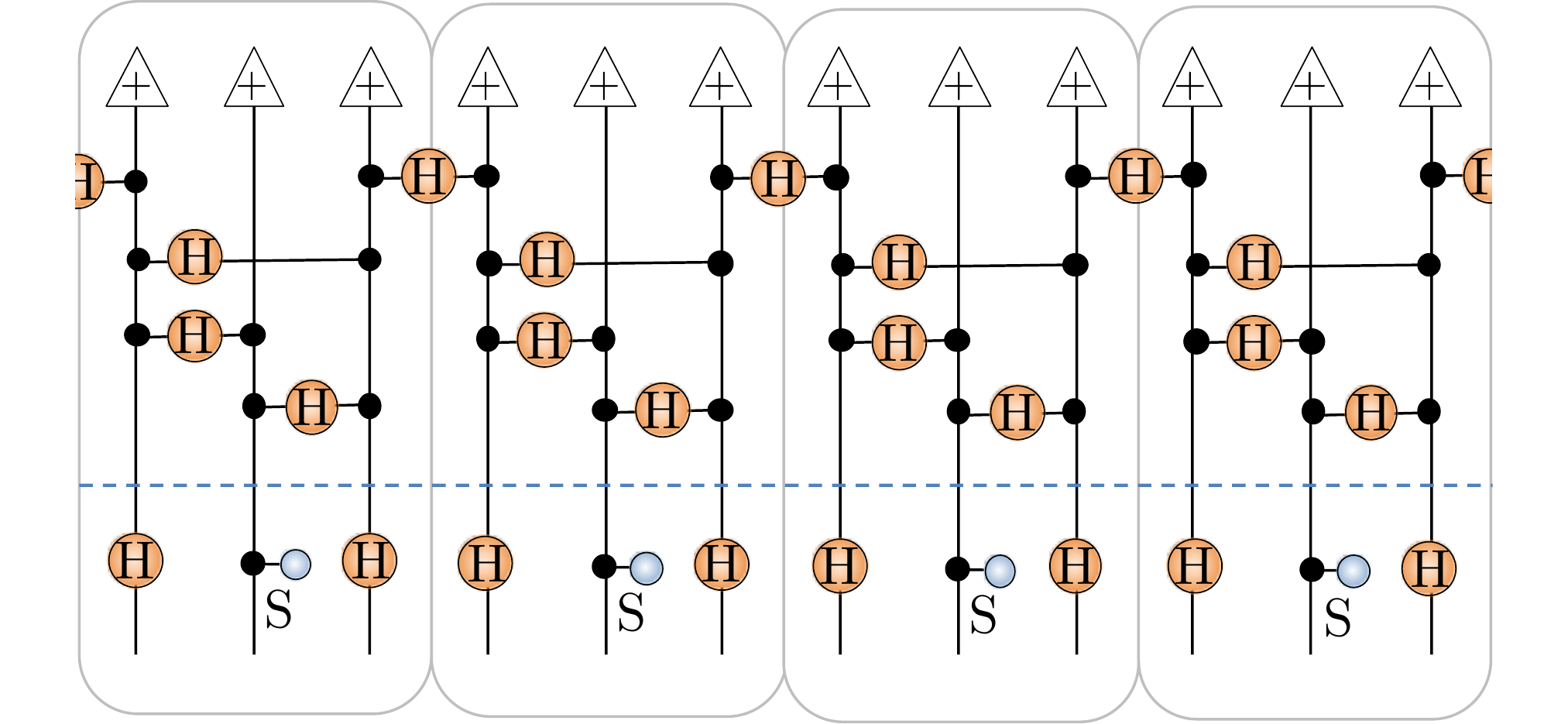} ,\label{eq:aklt_stabilizer}
\end{equation}
where the top part (above the dashed line) generates a graph state, and the bottom part applies local Clifford gates. Once unary projection to the spin-1 is applied $\ket{\Psi_{\rm unary}}$ generates the AKLT state~\cite{lu19}.

Previously in Ref.~\cite{Pei21} we showed how any stabilizer state for $N$ spin-$\half$'s can be readily converted into an NQS with $M \leq N-1$. Here we just summarise the basic process. The first step in this conversion is to use the local Clifford equivalence of stabilizer states to graph states to relocate all the non-diagonal Clifford gates to independent vertices of the graph. This conversion takes the simple chain-like graph state and pattern of Clifford gates from \eqr{eq:aklt_stabilizer} and gives the following for $N=12$ spins:
\begin{equation*}
\includegraphics[scale=0.45,valign=c]{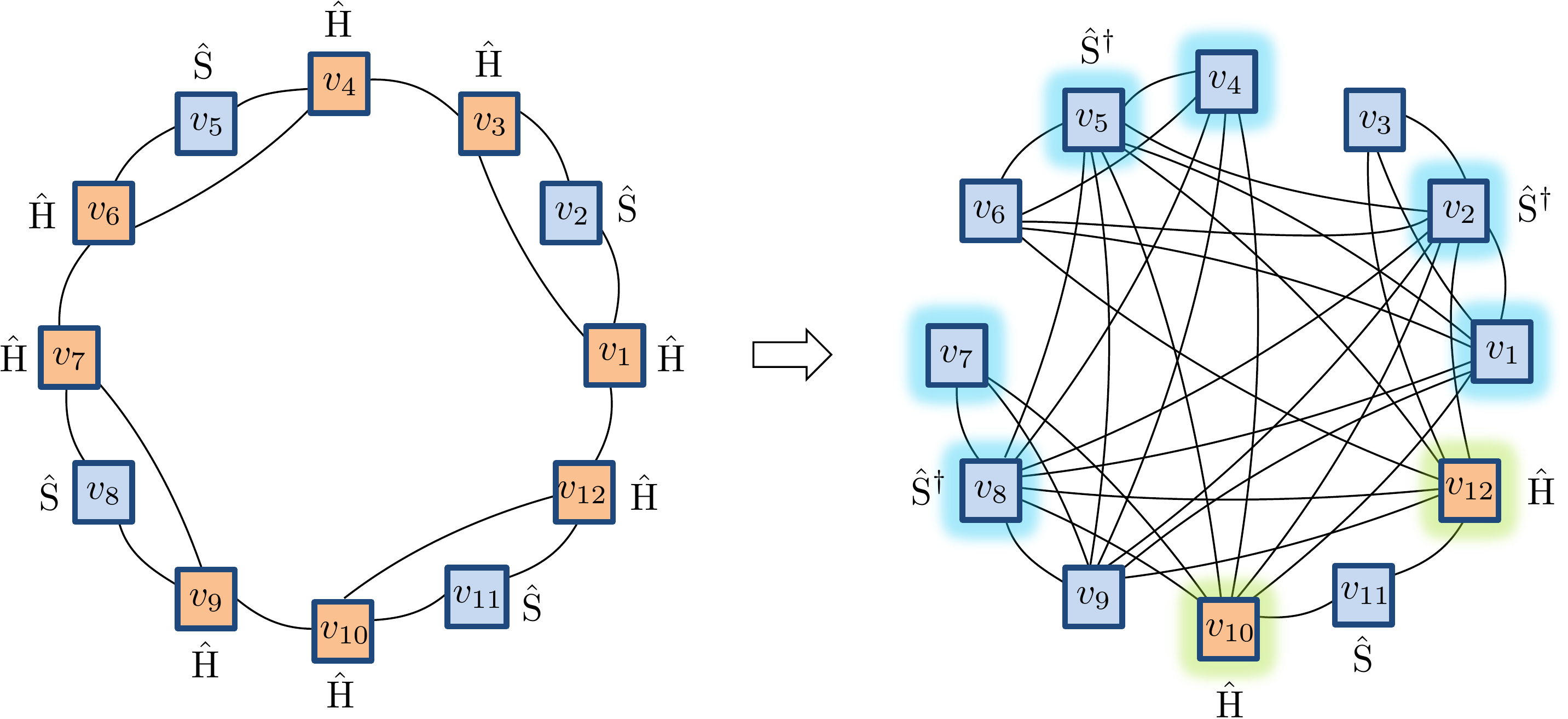} .
\end{equation*}
As required the resulting graph has diagonal Clifford gates on all vertices except for a small independent set $\mathcal{I} = \{10, 12\}$ highlighted. Notice that the three-site translational invariance of $\ket{\Psi_{\rm unary}}$, mirrored by the initial chain graph, is still formally present in the transformed graph but is now obscured by its highly connected topology. By forming a vertex cover $\mathcal{C} = \{\mathcal{I}, 1,2,4,5,7,8\}$ we obtain a NQS~\cite{Pei21} with $M=8$ hidden units\footnote{Despite the more complex graph this is the same number of hidden units required to describe the initial chain graph state as an NQS.}:
\begin{equation}
\quad\includegraphics[scale=0.5,valign=c]{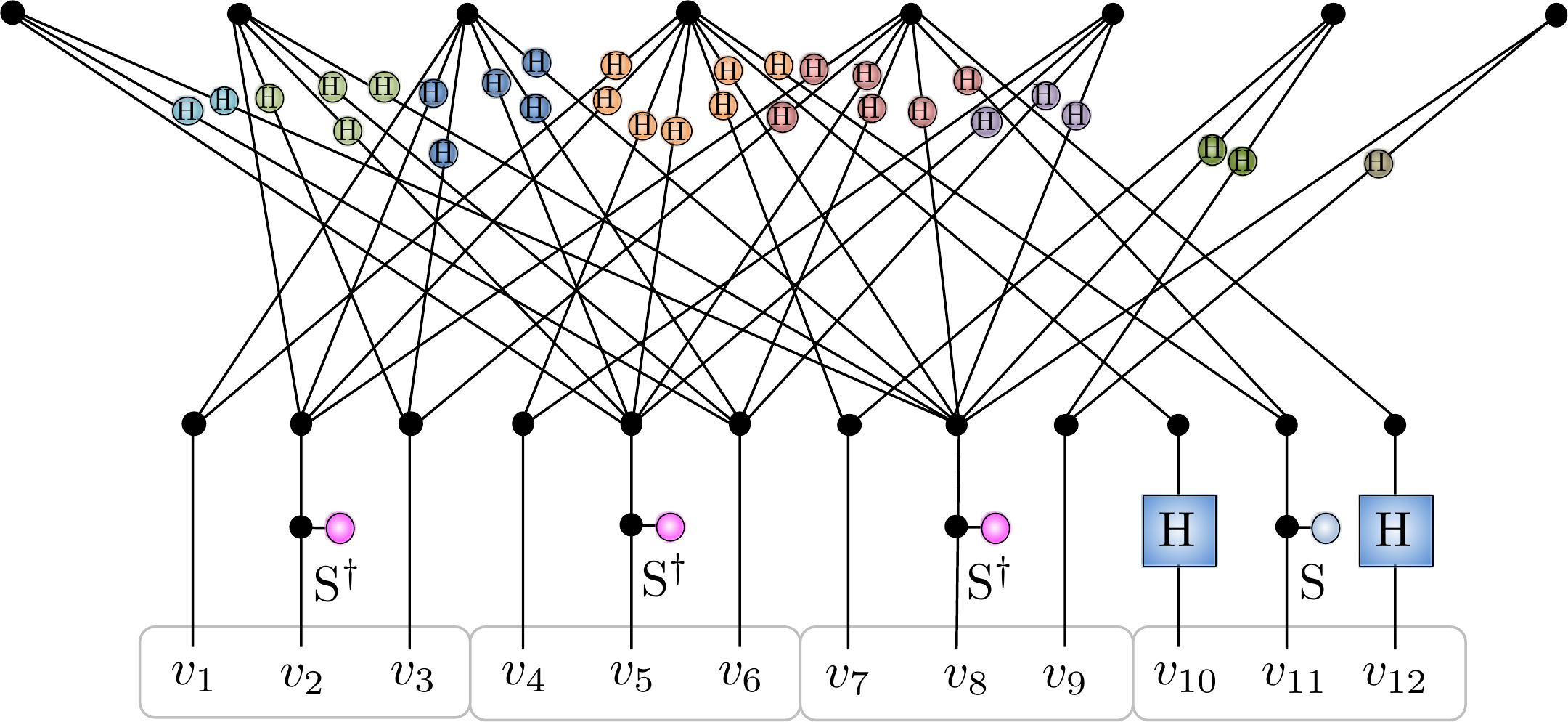} .
\end{equation}
More generally for $N$ spin-$\half$'s this procedure generates an NQS with $M=2N/3$ hidden units.

After applying the unary projection and contraction process directly to this spin-$\half$ NQS we obtain spin-1 NQS tensor network, whose schematic structure is shown here for $N=4$
\begin{equation}
\includegraphics[scale=0.5,valign=c]{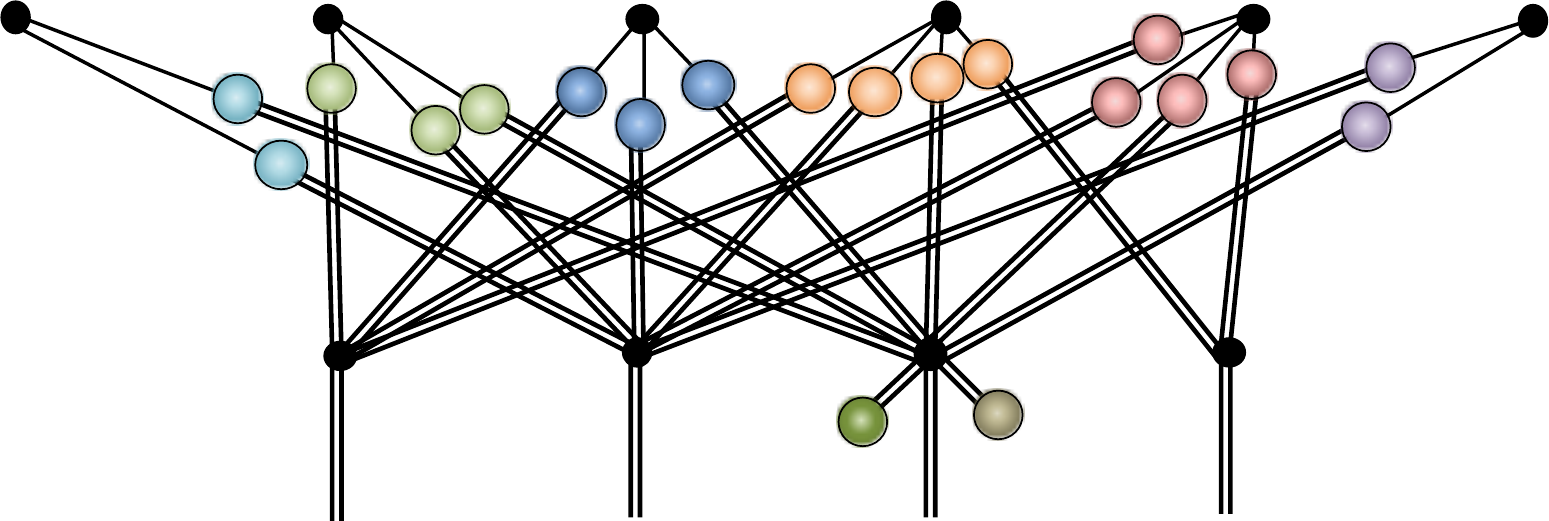} .
\end{equation}
It is evident that pairs of hidden units possess coordination $4,3$ and $2$, while one pair gets projected down to a spin-1 visible bias, giving $M=6$ overall. The same structure applies for general $N$ with pairs of equal coordination spanning $N,N-1,\dots,2$ giving a spin-1 NQS for the AKLT state in the $xyz$ basis requiring $M= 2N-2$ hidden units. This representation is essentially identical to the one presented in \secr{sec:aklt_xyz}, except that the hidden units implementing the nodal structure (the fully connected pair) also contribute to the sign structure, reducing the total hidden unit count by one pair. This raises an interesting question of whether an even more compressed spin-1 NQS representation of the AKLT state is possible. We finish by examining this using direct numerical optimisations.

\subsection{Numerical example - AKLT in $xyz$ and $S^{z}$ bases}\label{sec:aklt_numeric}
\begin{figure}[h]
\begin{center}
\includegraphics[scale=0.5]{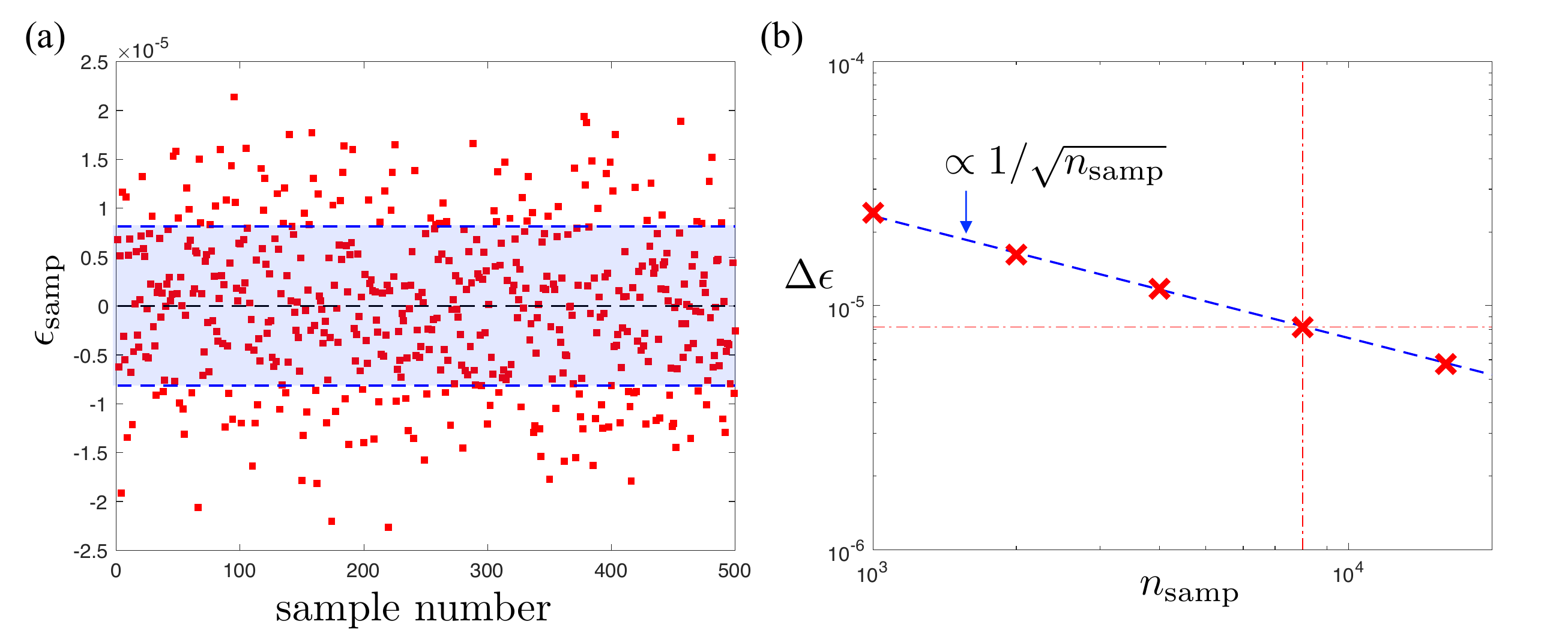}
\end{center}
\caption{(\textbf{a}) The energy $\epsilon_{\rm samp}$ (red squares) for 500 independent Monte Carlo runs of the $N=12$ exact AKLT spin-1 NQS each consisting of $n_{\rm samp} = 8000$ individual sampling steps separated by $N$ individual Markov chain moves to reduce autocorrelation effects. The standard deviation $\Delta\epsilon$ for these samples around the exact zero ground state energy is shown by the blue band. (\textbf{b}) The standard deviations $\Delta\epsilon$ of energies sampled versus the number of samples $n_{\rm samp}$ used. Each point is also calculated from 500 independent samples, and the number of sampling steps per evaluation $n_{\rm samp} = \{1000, 2000, 4000, 8000, 16000\}.$ The fluctuations in energy closely follow a $1/\sqrt{n_{\rm samp}}$ scaling (dashed line), as expected for a Monte Carlo sampling process~\cite{becca17}. The red dotted lines are to guide the eye to the point for $n_{\rm samp} = 8000$, which is representative of an optimisation step for our VMC calculations.}
\label{fig:aklt_samples}
\end{figure}

\begin{figure}[htb]
\begin{center}
\includegraphics[scale=0.5]{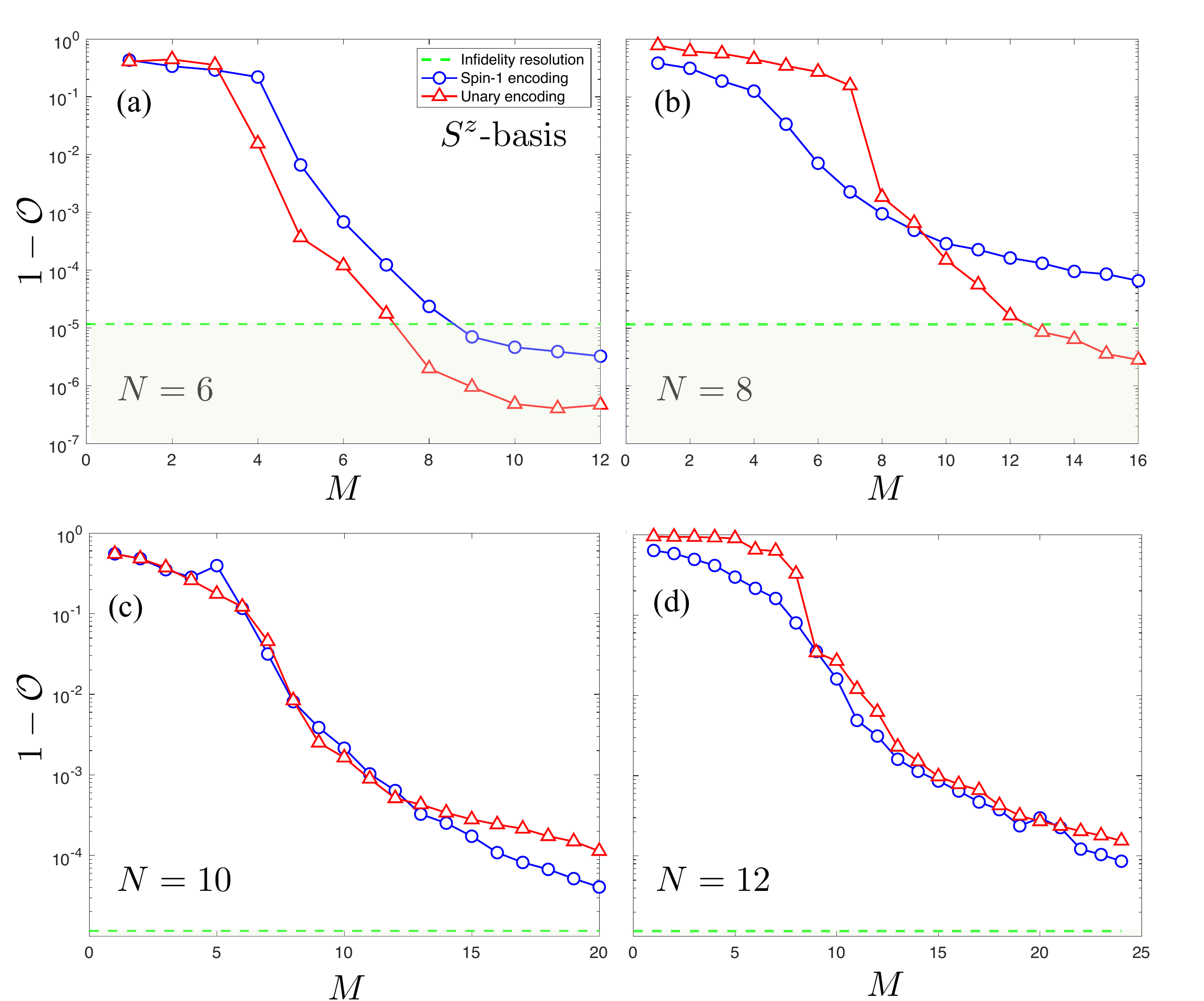}
\end{center}
\caption{Plots of the infidelity of unary (red stars) and spin-1 (blue squares) NQS with the AKLT state in the $S^{z}$ basis. The infidelity bound $\mathcal{R}$ is also plotted as a dotted green line. Results for four system sizes are plotted: (\textbf{a}) $N=6$, (\textbf{b}) $N=8$, (\textbf{c}) $N=10$, (\textbf{d}) $N=12$.}
\label{fig:aklt_sz_numeric}
\end{figure}

\begin{figure}[htb]
\begin{center}
\includegraphics[scale=0.5]{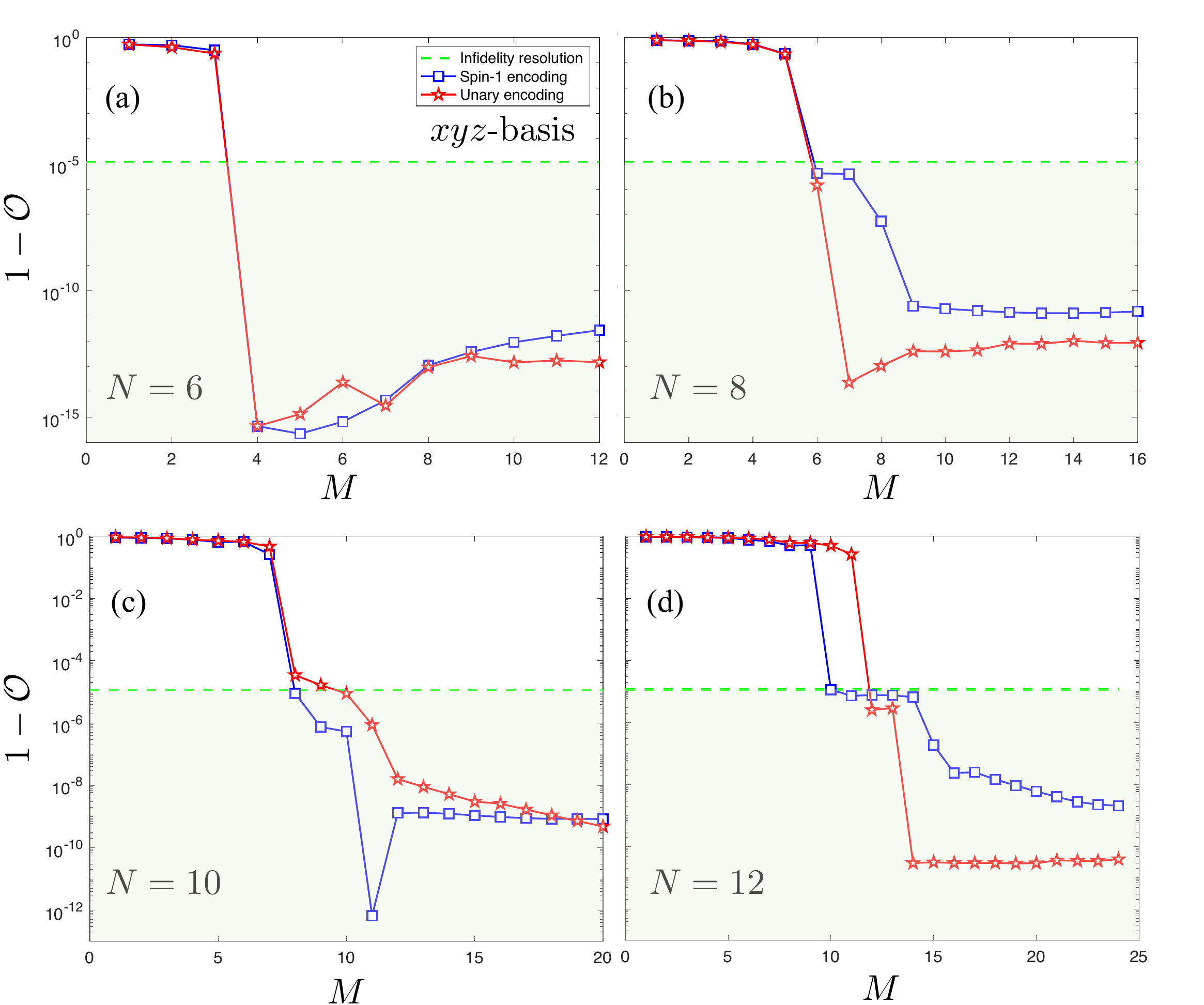}
\end{center}
\caption{Plots of the infidelity of unary (red stars) and spin-1 (blue squares) NQS with the AKLT state for the $xyz$ basis. The infidelity resolution $\mathcal{R}$ is plotted as a green dotted line. Results for four system sizes are plotted: (\textbf{a}) $N=6$, (\textbf{b}) $N=8$, (\textbf{c}) $N=10$, (\textbf{d}) $N=12$.}
\label{fig:aklt_xyz_numeric}
\end{figure}
Although the AKLT state is translationally invariant the hidden units encoding the sign structure of this solution are neither individually translationally invariant nor do their translates appear. Consequently we performed VMC calculations with increasing $M$ using both the spin-1 NQS and unary NQS for $N = \{6, 8, 10, 12 \}$ and compared against exact diagonalisation. As with the AFH model (which is $\hat{H}_{\rm AKLT}$ with $\beta = 0$) earlier in \secr{sec:afh} we considered both the $S^{z}$ and $xyz$ basis with their corresponding nodal structure enforced by sampling. We also utilise the same sequential growth scheme as we used in the Heisenberg calculations, again confirming the robustness of our qualitative conclusions by performing reruns of the optimisation sequence and presenting the best results here.
\par

As we are performing stochastic optimisation intrinsic sampling noise will limit the accuracy to which any formally exact solution can be found. It is therefore crucial to quantitatively characterise when exactness may have been reached numerically. For a gapped Hamiltonian the average energy of an approximate state $E$ can be related to its infidelity with the ground state using~\cite{becca17}
\begin{equation}
    1 - \mathcal{O} \leq \frac{E - E_{0}}{\delta} = \frac{\epsilon}{\delta} \label{eq:infidelity}
\end{equation}
where $\delta$ is the energy gap of the Hamiltonian and $\epsilon = E - E_{0}$ is the energy deviation. As the ground state energy $E_{0}$ of the AKLT Hamiltonian is zero, the energy deviation is simply the sampled energy of the state $\epsilon_{\rm samp}$. As shown in \fir{fig:aklt_samples}(a) even if the exact analytic spin-1 NQS solution is used $\epsilon_{\rm samp}$ fluctuates when using a finite number of samples typical of an optimisation step\footnote{For all optimisations presented in this paper, we used the following hyperparameters: number of samples per optimisation step $n_{\rm samp} = 8000$, number of optimisation steps $n_{\rm step} \sim 5000$. Typically the full wavefunction and its fidelity are calculated and checked every 1000 steps to gauge whether the solution has converged or requires further optimisation.}. To account for fluctuations caused by a finite $n_{\rm samp}$ employed throughout the stochastic optimisation we therefore use in \eqr{eq:infidelity} $\epsilon = \Delta\epsilon$, the standard deviation of the sampled energy. As shown in \fir{fig:aklt_samples}(b) $\Delta\epsilon$ vanishes as $1/\sqrt{n_{\rm samp}}$, and we estimate an algorithmic fidelity resolution of $\mathcal{R}  = \Delta\epsilon/\delta \approx 1.2 \times 10^{-5}$ for $N = 12$ sites below which it may be hard to discriminate an exact solution from an extremely good approximate one.
\par

In \fir{fig:aklt_sz_numeric} we show the smooth decrease in $1 - \mathcal{O}$ against $M$ for the $S^{z}$ basis. While the $N=6$ curve in \fir{fig:aklt_sz_numeric}(a) drops below $\mathcal{R}$ this is not attained for larger $N$ shown in \fir{fig:aklt_sz_numeric}(b)-(d) within the $M$'s tested. This indicates that no ``exact" NQS solutions have been located. As with the AFH model the spin-1 NQS and unary NQS achieve a similar accuracy verses $M$, although the former utilises less variational parameters. 
\par

The analogous results in \fir{fig:aklt_xyz_numeric} for the $xyz$ basis display remarkable features in comparison. For each $N$ a sharp drop in $1-\mathcal{O}$ by over 4 orders of magnitude is observed at $M=N-2$ that consistently pushes the infidelity below $\mathcal{R}$. After reaching this point, $1-\mathcal{O}$ versus $M$ plateaus, and subsequent hidden units have negligible bearing on the accuracy of the wavefunction due to statistical fluctuations originating from the stochastic optimisation. These features are consistently produced by both the spin-1 NQS and unary NQS, aside from the largest system size $N=12$ in \fir{fig:aklt_xyz_numeric}(d) indicating that the increasing number of redundant variational parameters in the unary encoding is complicating the optimisation. The overall behaviour of $1-\mathcal{O}$ observed in this basis is strong evidence that a numerically exact solution with $M = N$ (once the 2 hidden units implementing the nodal structure are included). This is substantially smaller than the analytic solutions introduced and is very suggestive of there being a {\em compact} exact spin-1 NQS representation of the AKLT state in the $xyz$ basis. 
\par

\section{Conclusion} \label{sec:conclusion}

We have introduced the most natural and direct generalisation of RBM from spin-$\half$ to spin-1. This necessitated including a quadratic visible bias and a quadratic visible-hidden interaction in the RBM energy function to ensure trivial product state representation, labelling freedom and gauge equivalence to the tensor network formulation. We demonstrated its use numerically for the spin-1 AFH model in both the $S^{z}$ and $xyz$ bases, illustrating how the choice of basis can affect the accuracy and hidden unit complexity of an NQS representation. Using our spin-1 NQS we then re-examined how to represent the AKLT state exactly. In the $S^z$ basis it is known to require $M \sim O(N^2)$ hidden units, yet by changing to the $xyz$ basis we construct an NQS with $M \sim O(N)$ hidden units. 
\par
Numerical VMC calculations have indicated that by capturing the nodal structure, either implicitly within the sampling or explicitly through the inclusion of extra hidden units, the optimisation can find even more efficient constructions for the sign structure. The resulting spin-1 NQS for the AKLT state in the $xyz$ basis requires $M = N$ hidden units in total making it {\em compact}. This example raises the interesting possibility of improving the efficiency and accuracy of NQS calculations by including single-spin basis transformations to lower the hidden unit complexity. 
\par
Several important open questions remain about NQS representations. In particular it would be instructive to build representations of classes of bosonic states using multinomial RBMs. In this case a local Fock basis is typically employed, however our findings suggest that it could be useful to explore a local basis that breaks the particle number symmetry when describing condensates. Moreover, the elevation of visible units from binary to multinomial raises the question of whether also using multinomial hidden units can enhance the expressiveness of NQS. This has been explored in the context of binary visible units in Ref.~\cite{rrapaj2021} in an analytical context to precisely represent certain two- and three-body interactions. The use of multinomial hidden units for numerical VMC calculations has been largely unexplored and is the subject of forthcoming work~\cite{Pei21_bhm} for the Bose Hubbard model in two dimensions.
\par

\authorcontributions{Conceptualization: M.P. and S.C.; methodology, M.P. and S.C.; software, M.P. and S.C; validation, M.P. and S.C.; formal analysis, M.P. and S.C.; investigation, M.P.; resources, S.C.; data curation, M.P.; writing---original draft preparation, S.C.; writing---review and editing, M.P.; visualization, M.P. and S.C.; supervision, S.C.; project administration, S.C.; funding acquisition, S.C. All authors have read and agreed to the published version of the manuscript.}

\funding{This research was funded by the Engineering and Physical Sciences Research Council (EPSRC) under grant No. EP/P025110/2. MP also acknowledges the University of Bristol Advanced Computing Research Centre for the use of their High Performance Computing facility (BlueCrystal Phase 3 and 4) in performing the VMC calculations presented.}

\dataavailability{{\tt MATLAB} scripts and .mat files containing the data shown in the figures are available in the data repository Ref.~\cite{nqs_s1_data}, doi:10.5523/bris.1ln9kyt6i86n12ehhftht27edp} 

\informedconsent{Not applicable.}

\conflictsofinterest{The authors declare no conflict of interest. The funders had no role in the design of the study; in the collection, analyses, or interpretation of data; in the writing of the manuscript, or in the decision to publish the results.}

\newpage

\appendixstart
\appendix

\section{Monte Carlo method and sampling} \label{sec:samp}

Any many-body quantum state can be expressed in terms of its amplitudes on a set of configurations states in a chosen basis, but exhaustively computing the amplitudes for every possible configuration in the basis is often prohibitively expensive. Monte Carlo methods circumvent this by randomly selecting a subset of these configurations, and calculating an estimate of key observables using this selection. In this case we require only that the (unnormalised) amplitudes $\Psi_{\bm p}({\bm S})$ of the ansatz (defined by parameters $\bm p$) in a fixed basis, such as $S^z$, can be computed in a time with polynomial complexity in $N$. We then rewrite the expectation value as
\begin{equation}
\av{\hat{A}}_{\Psi_{\bm p}} = \sum_{\bm S} P({\bm S}) A({\bm S}), \quad {\rm where} \quad P({\bm S}) =  \frac{|\Psi_{\bm p}({\bm S})|^2}{\sum_{\bm S} |\Psi_{\bm p}({\bm S})|^2},
\end{equation}
is the probability of the spin configuration $\bm S$ and
\begin{equation}
A({\bm S}) = \sum_{{\bm S}'} \frac{\Psi_{\bm p}({\bm S}')}{\Psi_{\bm p}({\bm S})}\bra{\bm S}\hat{A}\ket{{\bm S}'}, \label{eq:estimator}
\end{equation}
is the local estimator of $\hat{A}$. The sum over ${\bm S}'$ in \eqr{eq:estimator} is restricted to only those configurations for which the matrix element $\bra{\bm S}\hat{A}\ket{{\bm S}'} \neq 0$, and so as long as $\hat{A}$ is sparse in the chosen fixed basis then $A({\bm S})$ can be efficiently evaluated. Typical product terms comprising short-ranged Hamiltonians $\hat{H}$ fulfil this requirement. Thus an unbiased estimate of $\av{\hat{A}}_{\Psi_{\bm p}}$ is made by drawing independent samples of spin configurations $\bm S$ from the distribution $P({\bm S})$. A powerful way to accomplish this is Markov-chain Monte Carlo in which a sequence of spin configurations ${\bm S}_0 \rightarrow {\bm S}_1 \rightarrow \cdots$ is generated using the Metropolis-Hastings algorithm that accepts a proposed configuration ${\bm S}_{\rm prop}$ for the $n$th iteration with a probability $P_n({\bm S}_{\rm prop}) = {\rm min}[1,|\Psi_{\bm p}({\bm S}_{\rm prop})|^2/|\Psi_{\bm p}({\bm S}_{n-1})|^2]$. A subset of visited configurations in the sequence, suitably separated to ensure they are decorrelated, is then used to estimate of $\av{\hat{A}}_{\Psi_{\bm p}}$. Crucially both $A({\bm S})$ and Monte Carlo sampling rely only on the ratio of ansatz amplitudes, so the intractable state normalisation $\sum_{\bm S} |\Psi_{\bm p}({\bm S})|^2$ is entirely avoided. 
\par
Variational Monte Carlo proceeds by evaluating the ansatz energy $E_{\bm p} = \av{\hat{H}}_{\Psi_{\bm p}}$ and its variance, along with their gradient vectors with respects to parameters $\bm p$. The parameters can then be updated by a small step along the direction of steepest descent, with this iterated until convergence to the minimum ${\bm p}_0$~\cite{gubernatis16}. The efficiency of this minimisation process is strongly problem and ansatz dependent. In challenging cases more sophisticated approaches such as modified stochastic optimisation~\cite{lou07}, the `linear method'~\cite{nightingale01,toulouse07} and stochastic reconfiguration~\cite{sorella01,becca17} are needed. All numerical calculations in this paper employ stochastic reconfiguration. The method involves the construction of a $n_{\rm params} \times n_{\rm params}$ matrix, requiring $O(n_{\rm samp}n_{\rm params}^{2})$ operations, though with a conjugate gradient method this can be brought down to $O(n_{\rm samp}n_{\rm params})$~\cite{neuscamman12}. The parameter changes can then be calculated from a set of linear equations involving the matrix, which scales as $O(n_{\rm params}^2)$ using matrix-vector product methods. As a rule of thumb the matrix requires a number of samples $n_{\rm samp} > 10 n_{\rm params}$ to ensure the sampled matrix is not rank-deficient~\cite{becca17}, giving an overall minimum scaling of $O(n_{\rm params}^2)$ for the algorithm, though typically it is prudent to have a large number of samples $n_{\rm samp} \gg n_{\rm params}$ as the error of any sampled observable typically scales as $1/\sqrt{n_{\rm samp}}$~\cite{becca17}, including the elements of the matrix.

\section{Boltzmann parameterisation of coupling matrices} \label{app:boltzmann}

Every $2 \times 3$ coupling matrix within the spin-1 NQS tensor network can be parameterised in Boltzmann form using the generalised energy function introduced in \secr{sec:spin1rbm} as
\begin{equation}
C^{(ij)}_{h_iv_j} = \exp\left(\tilde{c}_{ij} + w_{ij}h_iv_j + W_{ij}h_iv_j^2 + \tilde{b}_{ij}h_i + \tilde{a_{ij}}v_j + \tilde{A}_{ij}v_j^2\right), \label{eq:gen_rbm_coupling_mat}
\end{equation}
for $h_i \in \{+1,-1\}$ and $v_j \in \{+1,0,-1\}$, which includes a quadratic weight $W_{ij}$ and quadratic partial bias $\tilde{A}_{ij}$. These complex parameters are found from the coupling matrix elements as
\begin{eqnarray*}
\tilde{a}_{ij} = \frac{1}{4}\left[\log\left(C^{(ij)}_{++}\right) + \log\left(C^{(ij)}_{-+}\right) - \log\left(C^{(ij)}_{+-}\right) - \log\left(C^{(ij)}_{--}\right)\right], \\
\tilde{b}_{ij} = \frac{1}{2}\left[\log\left(C^{(ij)}_{+0}\right) - \log\left(C^{(ij)}_{-0}\right)\right], \\
\tilde{c}_{ij} = \frac{1}{2}\left[\log\left(C^{(ij)}_{+0}\right) + \log\left(C^{(ij)}_{-0}\right)\right], \\
w_{ij} = \frac{1}{4}\left[\log\left(C^{(ij)}_{++}\right) - \log\left(C^{(ij)}_{-+}\right) - \log\left(C^{(ij)}_{+-}\right) + \log\left(C^{(ij)}_{--}\right)\right], \\
W_{ij} = \frac{1}{4}\left[\log\left(C^{(ij)}_{++}\right) - \log\left(C^{(ij)}_{-+}\right) - 2\log\left(C^{(ij)}_{+0}\right) + 2\log\left(C^{(ij)}_{-0}\right) + \log\left(C^{(ij)}_{+-}\right) - \log\left(C^{(ij)}_{--}\right)\right], \\
\tilde{A}_{ij} = \frac{1}{4}\left[\log\left(C^{(ij)}_{++}\right) + \log\left(C^{(ij)}_{-+}\right) - 2\log\left(C^{(ij)}_{+0}\right) - 2\log\left(C^{(ij)}_{-0}\right) + \log\left(C^{(ij)}_{+-}\right) + \log\left(C^{(ij)}_{--}\right)\right]. 
\end{eqnarray*}
In both cases the presence of zero matrix elements in ${\bf C}^{(ij)}$, which appear frequently in our `by hand' constructions, return diverging parameters. However, as discussed in Ref.~\cite{Pei21} this can be handled by softening the zeros via 
\begin{equation}
C^{(ij)}_{h_iv_j} \mapsto \max\left(C^{(ij)}_{h_iv_j},e^{-\mathcal{S}}\right),
\end{equation}
where typically $\mathcal{S} \approx 5-10$.
\par

\section{AKLT state NQS in $S^{z}$ basis} \label{app:aklt_sz}
In this appendix we construct a spin-1 NQS for the AKLT state amplitudes in the $S^z$ basis given in \eqr{eq:aklt_sz_amps} by using hidden units as successive filters. The rules governing the structure of the amplitudes can be summarised as follows:
\begin{enumerate}
\item Zero out the amplitude for any configuration containing a substring for any $\ell=0,1,\dots,N-2$ of the form $+[0 \cdots 0]^\ell+$ or $-[0\cdots 0]^\ell-$, where $[0\cdots 0]^\ell$ is a string of 0's of length $\ell$. To implement this rule we introduce the following $2 \times 3$ coupling matrices
\begin{equation*}
\qquad{\bf C}_{{\rm i}\Uparrow} = \left[
\begin{array}{ccc}
1 & 1 & 1  \\
{\rm i}  & 0 & 0   
\end{array}
\right], \quad {\bf C}_{{\rm i}\Downarrow} = \left[
\begin{array}{ccc}
1 & 1 & 1  \\
0 & 0 & {\rm i}
\end{array}
\right], \quad {\bf C}_{+0} = \left[
\begin{array}{ccc}
1 & 1 & 1  \\
0 & 1 & 0
\end{array}
\right],
\end{equation*}
which in turn single out $S^z$ basis spin-1 states $\ket{\Uparrow}$, $\ket{0}$ and $\ket{\Downarrow}$. To cancel out in any configuration substrings $+[0 \cdots 0]^\ell+$ and $-[0 \cdots 0]^\ell-$, starting at a given site $k$ with a given separation $\ell$, we construct pairs of hidden units with 
\begin{eqnarray*}
\{{\bf I}, \cdots, {\bf I}, \overbrace{{\bf C}_{+{\rm i}}}^{k}, {\bf C}_{+0}, \cdots, {\bf C}_{+0},\overbrace{{\bf C}_{+{\rm i}}}^{k+1+\ell},{\bf I}, \cdots, {\bf I}\}, \quad {\rm and}\\ 
\{{\bf I}, \cdots, {\bf I}, \overbrace{{\bf C}_{-{\rm i}}}^{k}, {\bf C}_{+0}, \cdots, {\bf C}_{+0},\overbrace{{\bf C}_{-{\rm i}}}^{k+1+\ell},{\bf I}, \cdots, {\bf I}\}.
\end{eqnarray*}
To capture all separations $\ell = 0,1,\dots,N-2$ starting on any site $k=1,2,\dots,N$ we require $2N(N-1)$ hidden units.
\item Apply a factor of $(-1)^\ell$ to the amplitude of a configuration for each substring of the form $+[0\cdots 0]^\ell-$ it contains. To implement this rule we introduce the two additional coupling matrices
\begin{equation*}
{\bf C}_{2\Uparrow} = \left[
\begin{array}{ccc}
1 & 1 & 1  \\
2 & 0 & 0   
\end{array}
\right], \quad {\bf C}_{-\Downarrow} = \left[
\begin{array}{ccr}
1 & 1 & 1  \\
0 & 0 & -1
\end{array}
\right].
\end{equation*}
A factor $(-1)^\ell$ is induced on any configuration containing the substring $+[0\cdots 0]^\ell-$, starting at a given site $k$ with a given separation $\ell$, by a hidden unit with 
\begin{equation*}
\{{\bf I}, \cdots, {\bf I}, \overbrace{{\bf C}_{2\Uparrow}}^{k}, {\bf C}_{+0}, \cdots, {\bf C}_{+0},\overbrace{{\bf C}_{-\Downarrow}}^{k+1+\ell},{\bf I}, \cdots, {\bf I}\}.
\end{equation*}
To capture all odd separations $\ell = 1,3,\dots,N-2$ starting on any site $k=1,2,\dots,N$ requires $N \lfloor\half(N-1)\rfloor$ hidden units.
\item Zero out the single excitation configurations $+[0\cdots 0]^{N-1}$ and $-[0\cdots 0]^{N-1}$ and all their translates. To implement this rule we introduce another coupling matrix
\begin{equation*}
{\bf C}_{-\Uparrow} = \left[
\begin{array}{rcc}
1 & 1 & 1  \\
-1 & 0 & 0   
\end{array}
\right].
\end{equation*}
Configurations $+[0\cdots 0]^{N-1}$ and $-[0\cdots 0]^{N-1}$ are then cancelled out by a pair of hidden units with $\{{\bf C}_{-\Uparrow}, {\bf C}_{+0}, \cdots, {\bf C}_{+0}\}$ and $\{{\bf C}_{-\Downarrow}, {\bf C}_{+0}, \cdots, {\bf C}_{+0}\}$. Capturing all the translates is achieved with $2N$ hidden units.
\item For $N$ odd zero out the amplitude for configuration $[0\cdots 0]^N$, or for $N$ even double its amplitude. To implement this rule we introduce one final coupling matrix
\begin{equation*}
{\bf C}_{-0} = \left[
\begin{array}{crc}
1 & 1 & 1  \\
0 & -1 & 0   
\end{array}
\right],
\end{equation*}
and defining a single hidden unit as $\{{\bf C}_{-0}, {\bf C}_{-0}, \cdots, {\bf C}_{-0}\}$.
\item Apply a factors of $2^{(\#~+{\rm 's})}\times (-1)^{(\#~0{\rm 's})} \times (-1)^{(\#~-{\rm 's})}$ to the amplitudes of all configurations. This does not require any additional hidden units. Instead we can pick any hidden unit and right multiply each of its $N$ coupling matrices by ${\bf D} = {\rm diag}(2,-1,-1)$. In RBM formalism this is equivalent to setting non-zero visible unit biases.
\end{enumerate}
This gives a total of $M = 2N^2 + N \lfloor\half(N-1)\rfloor  + 1$ hidden units. Notice that the origin of the $O(N^2)$ scaling arises from spatial translations, so if this symmetry is directly imposed within the NQS ansatz, as it often is numerically, then the AKLT state in the $S^z$ basis instead has its number of unique hidden units scaling as $M = 2N + \lfloor\half(N-1)\rfloor + 1$.
\par

\end{paracol}

\section*{References}
\externalbibliography{yes}
\bibliography{nqs_graph}

\end{document}